\documentclass[sigconf]{acmart}

\copyrightyear{2026}
\acmYear{2026}
\setcopyright{cc}
\setcctype{by}
\acmConference
    [CIKM '26]
    {Proceedings of the 35th ACM International Conference on Information and Knowledge Management}
    {November 07--11, 2026}
    {Rome, Italy}
\acmBooktitle
    {Proceedings of the 35th ACM International Conference on Information and Knowledge Management (CIKM '26), November 07--11, 2026, Rome, Italy}
\acmDOI{10.1145/3799682.3841118}
\acmISBN{979-8-4007-2539-5/2026/11}

\usepackage{enumitem}
\usepackage{amsmath}
\usepackage{multirow}
\usepackage{subcaption}
\usepackage{hhline}

\theoremstyle{definition}
\newtheorem{problem}{Problem}

\begin{document}

\title
    [TRACER: Balancing Stability-Plasticity-Cognitivity Trilemma for LLM Enhanced Continual Recommendation]
    {TRACER: Balancing Stability-Plasticity-Cognitivity Trilemma\\for LLM Enhanced Continual Recommendation}

\author{WooJoo Kim}
\email{kimuj0103@postech.ac.kr}
\affiliation
{
    \institution{Pohang University of\\Science and Technology}
    \city{Pohang}
    \country{Republic of Korea}
}
\author{HyunSik Yoo}
\email{hy40@illinois.edu}
\affiliation
{
    \institution{University of Illinois\\Urbana-Champaign}
    \city{Urbana}
    \state{IL}
    \country{USA}
}
\author{JunYoung Kim}
\email{junyoungkim@postech.ac.kr}
\affiliation
{
    \institution{Pohang University of\\Science and Technology}
    \city{Pohang}
    \country{Republic of Korea}
}
\author{JaeHyung Lim}
\email{jaehyunglim@postech.ac.kr}
\affiliation
{
    \institution{Pohang University of\\Science and Technology}
    \city{Pohang}
    \country{Republic of Korea}
}
\author{SeongKu Kang}
\authornote{Corresponding authors.}
\email{seongkukang@korea.ac.kr}
\affiliation
{
    \institution{Korea University}
    \city{Seoul}
    \country{Republic of Korea}
}
\author{HwanJo Yu}
\authornotemark[1]
\email{hwanjoyu@postech.ac.kr}
\affiliation
{
    \institution{Pohang University of\\Science and Technology}
    \city{Pohang}
    \country{Republic of Korea}
}
\renewcommand{\shortauthors}{WooJoo Kim et al.}

\begin{abstract}
    Continual recommendation aims to capture evolving user interests from streaming data but struggles with sparsity.
    LLM enhancers mitigate this with semantic knowledge, but naive integration creates a new conflict.
    We identify this as the \textbf{Stability-Plasticity-Cognitivity (SPC) Trilemma}, where generalized LLM semantic priors (Cognitivity) conflict with retaining personalized historical preferences (Stability) and adapting to individual interest shifts (Plasticity).
    To address this, we propose \textbf{T}rilemma-\textbf{R}esponsive \textbf{A}daptive \textbf{C}ontinual \textbf{E}nhancement for \textbf{R}ecommendation (\textbf{TRACER}).
    TRACER synergistically combines three specialized modules, each targeting stability, plasticity, or cognitivity, while preventing any single lemma from dominating.
    This holistic design enables semantic knowledge to support history retention and adaptation to evolving interests without disrupting continual learning.
    Across five real-world datasets, TRACER effectively harmonizes the SPC trilemma and outperforms state-of-the-art baselines by up to 14.38\%.\footnote{Our code is available at \url{https://github.com/woo-joo/TRACER_CIKM26}.}
\end{abstract}



\maketitle

\section{Introduction}

Continual recommendation (CR) \cite{lee2024continual, lim2025federated, yoo2025continual} captures evolving user interests from streaming data but struggles with data sparsity \cite{zhang2024latent, choi2025dynamic, cui2023event, chen2021incremental}, as interactions at each time stage are often insufficient for effective ID embedding updates.
LLMs offer a remedy, leveraging rich knowledge and cognitive capabilities to perceive semantics beyond sparse identifiers \cite{sanner2023large, hou2024large, wu2024survey}.
However, directly employing LLMs as recommenders \cite{li2023prompt, kim2024large, bao2023tallrec, bao2025bi, lee2026filling} is impractical in CR due to prohibitive computational costs and inference latency that hinder rapid model updates in streaming environments.
Thus, a critical challenge remains: \textit{how to infuse LLM semantic understanding into CR systems without compromising efficiency?}

To address this, the LLM enhancer paradigm \cite{harte2023leveraging, hu2024enhancing, liu2024llm} augments ID-based recommenders with pre-computed LLM-derived semantic representations, avoiding direct LLM inference during training and serving.
Recent LLM enhancers go beyond merely encoding raw textual metadata of users or items.
They leverage LLM world knowledge and reasoning to transform such metadata (e.g., reviews and descriptions) into semantically coherent profiles such as user personas and item intents \cite{ren2024representation, xi2024towards}, which are then encoded into semantic representations.
This paradigm enables CR models to exploit rich semantic knowledge while preserving the efficiency of ID-based recommenders for rapid model updates.

However, naive adoption of LLM enhancers disrupts the balance between \textbf{Stability} (preserving historical preferences) and \textbf{Plasticity} (adapting to interest shifts), a core continual-learning challenge \cite{kirkpatrick2017overcoming, prabhu2020gdumb, lu2025rethinking}.
In LLM-enhanced CR, this balance is further shaped by how the model exploits semantic priors to inject the cognitive strengths of LLMs into recommendation.
We call this ability \textbf{Cognitivity}: leveraging semantic priors grounded in LLM world knowledge and reasoning to infer user-item relevance.
However, user preferences are individual and evolving, whereas generalized LLM semantic priors may diverge from collaborative signals.
Consequently, blindly pursuing cognitivity can erode core historical preferences or neglect genuine interest shifts, creating the \textbf{Stability-Plasticity-Cognitivity (SPC) Trilemma}.

\begin{figure*}[t]
  \centering
  \begin{subfigure}[t]{0.45\textwidth}
    \centering
    \includegraphics[width=\textwidth]{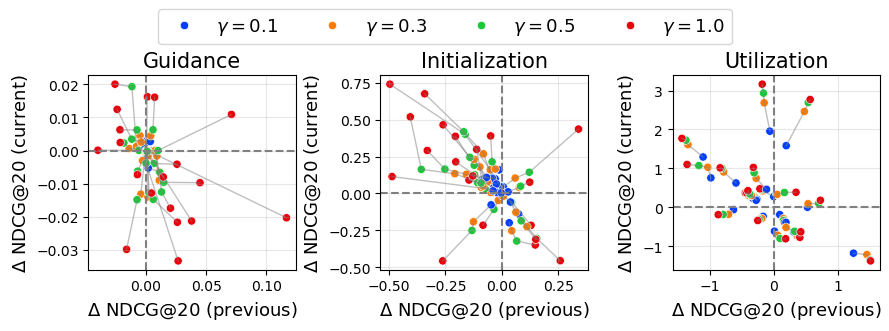}
    \caption{Stability-Plasticity Landscape}
  \end{subfigure}
  \hfill
  \begin{subfigure}[t]{0.15\textwidth}
    \centering
    \includegraphics[width=\textwidth]{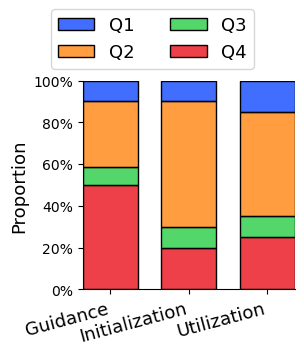}
    \caption{Distribution}
  \end{subfigure}
  \hfill
  \begin{subfigure}[t]{0.34\textwidth}
    \centering
    \includegraphics[width=\textwidth]{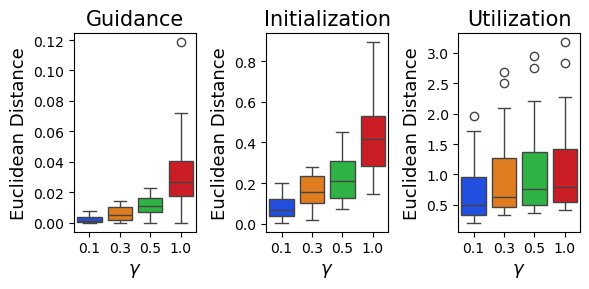}
    \caption{Euclidean Distance from Origin}
  \end{subfigure}
  \caption{Trilemma validation. (a) Performance deviations against plain LightGCN. Gray lines connect points from the same data block. (b) Distribution of points across the four quadrants. (c) Average Euclidean distance of points from the origin.}
  \label{fig/trilemma_validation}
  \Description{}
\end{figure*}

These trilemma-induced distortions manifest diversely.
First, with heterogeneous preferences (e.g., `classical' and `metal'), cognitivity may mistake semantic disparity for inconsistency.
Dismissing one side as noise obscures multifaceted preferences and suppresses recommendations from that side (stability$\downarrow$, plasticity$\downarrow$).
Second, when a `zero-sugar' consumer shifts to desserts, semantic priors may lock them into a `health-conscious' profile, enforcing past habits while blocking new interests (stability$\uparrow$, plasticity$\downarrow$).
Conversely, cognitivity can overreact to a coarse semantic cue, broadly mapping a `foam roller' purchased for `bodybuilding' to `wellness' and producing generic recommendations that capture recent signals but dilute fine-grained preferences (stability$\downarrow$, plasticity$\uparrow$).

To validate this trilemma, we construct three LLM enhancers using semantic representations in the training objective (\textit{Guidance}), parameter instantiation (\textit{Initialization}), and inference (\textit{Utilization}).
Each is integrated into LightGCN \cite{he2020lightgcn} for continual fine-tuning on chronological data blocks, with $\gamma \in [0,1]$ controlling LLM intervention intensity (see App.~\ref{sec/trilemma_validation_experiment}).
\autoref{fig/trilemma_validation}a shows NDCG@20 deviations from plain LightGCN on previous (stability) and current (plasticity) blocks.
Points cluster in quadrants II (stability$\downarrow$, plasticity$\uparrow$) and IV (stability$\uparrow$, plasticity$\downarrow$) (\autoref{fig/trilemma_validation}b), showing cognitivity improves stability or plasticity but degrades the other.
Higher $\gamma$ moves points farther from the origin (\autoref{fig/trilemma_validation}c), indicating stronger LLM intervention exacerbates this imbalance.
These results verify the SPC trilemma, motivating a dedicated solution for LLM-enhanced CR.

We further show that \textit{Guidance}, \textit{Initialization}, and \textit{Utilization} respectively bias LLM-enhanced CR toward stability, plasticity, and cognitivity (see Sec.~\ref{sec/spc_trilemma_analysis}).
Building on this, we propose \textbf{T}rilemma-\textbf{R}esponsive \textbf{A}daptive \textbf{C}ontinual \textbf{E}nhancement for \textbf{R}ecommendation (\textbf{TRACER}).
TRACER synergistically combines three redesigned enhancers: \textit{Semantic Synchronization} for guidance-driven stability, \textit{Procrustes Projection} for initialization-driven plasticity, and \textit{Confidence-gated Condensation} for utilization-driven cognitivity.
Each preserves its target-lemma strength while mitigating its bias, allowing semantic knowledge to support history retention and adaptation to evolving interests.
To our knowledge, this is the first framework to systematically integrate LLM enhancers into CR.

Our main contributions are as follows:
\begin{itemize}[itemsep=1pt, left=0pt]
    \item \textbf{SPC Trilemma}: We formally define the SPC trilemma and verify that blindly pursuing cognitivity induces unbalanced trade-offs.
    \item \textbf{LLM Enhancer Analysis}: We analyze how semantic representations induce distinct SPC biases across the model workflow.
    \item \textbf{TRACER Framework}: We propose TRACER for LLM-enhanced CR by synergistically combining three modules for SPC balance.
    \item \textbf{SOTA Performance}: Across five datasets, TRACER outperforms state-of-the-art baselines by harmonizing the SPC trilemma.
\end{itemize}

\section{Preliminary and Related Work}

\begin{table}[t]
    \scriptsize
    \setlength{\tabcolsep}{2pt}
    \caption{Key notations used in this paper.}
    \label{tab/notation}
    \begin{tabular}{c|l}
        \toprule
        \textbf{Notation} & \multicolumn{1}{c}{\textbf{Description}} \\
        \midrule
        $T$                                                                                    & total number of continual time stages \\
        $\mathcal{D}^{(t)}$                                                                    & interaction set at stage $t$ \\
        $\mathcal{U}^{(t)}$ / $\mathcal{I}^{(t)}$                                              & set of active users / items at stage $t$ \\
        $\mathcal{U}^{(t)}_\mathrm{new}$ / $\mathcal{I}^{(t)}_\mathrm{new}$                    & set of new users / items appearing first in $\mathcal{D}^{(t)}$ \\
        $\mathcal{U}^{(t)}_\mathrm{ext}$ / $\mathcal{I}^{(t)}_\mathrm{ext}$                    & set of existing users / items at stage $t$, i.e., $\mathcal{U}^{(t)} \setminus \mathcal{U}^{(t)}_\mathrm{new}$ / $\mathcal{I}^{(t)} \setminus \mathcal{I}^{(t)}_\mathrm{new}$ \\
        \midrule
        $\mathbf{e}^{(t)}_u$ / $\mathbf{e}^{(t)}_i$                                            & ID embedding of user $u$ / item $i$ at stage $t$ \\
        $\mathbf{E}^{(t)}$ / $\mathbf{E}^{(t)}_\mathrm{new}$ / $\mathbf{E}^{(t)}_\mathrm{ext}$ & ID embedding table for all / new / existing users and items at stage $t$ \\
        $\mathbf{x}_u$ / $\mathbf{x}_i$                                                        & semantic representation of user $u$ / item $i$ \\
        $\mathbf{X}$                                                                           & semantic representation table \\
        $\phi^{(t)}(\cdot)$                                                                    & global adapter for semantic representations at stage $t$ \\
        \midrule
        $\hat{y}_{ui}$                                                                         & predicted interaction score between user $u$ and item $i$ \\
        $\mathcal{L}_\mathrm{rec}$                                                             & recommendation loss \\
        \bottomrule
    \end{tabular}
\end{table}

\autoref{tab/notation} summarizes the key notations.
Throughout the paper, we employ the tilde ($\tilde{\cdot}$) and hat ($\hat{\cdot}$) accents to explicitly denote the initialized and optimized states of any learnable parameter, respectively.

\subsection{Continual Recommendation}

We define the CR task where a model, pretrained on a base dataset $\mathcal{D}^{(0)}$, adapts to sequential data blocks $\{ \mathcal{D}^{(t)} \}_{t=1}^T$.
For active entities (i.e., users and items) $\mathcal{V}^{(t)} = \mathcal{U}^{(t)} \cup \mathcal{I}^{(t)}$ in $\mathcal{D}^{(t)}$, we partition $\mathcal{V}^{(t)}$ into new entities $\mathcal{V}^{(t)}_\mathrm{new} = \mathcal{U}^{(t)}_\mathrm{new} \cup \mathcal{I}^{(t)}_\mathrm{new}$ and existing entities $\mathcal{V}^{(t)}_\mathrm{ext} = \mathcal{U}^{(t)}_\mathrm{ext} \cup \mathcal{I}^{(t)}_\mathrm{ext}$.
The model parameters $\theta^{(t)}$ include the ID embedding table $\mathbf{E}^{(t)} \in \mathbb{R}^{|\mathcal{V}^{(t)}| \times d}$, divided into $\mathbf{E}^{(t)}_\mathrm{new}$ and $\mathbf{E}^{(t)}_\mathrm{ext}$ for new and existing entities, respectively.
At each stage $t$, the goal is to update $\theta^{(t)}$ from an initialized state $\tilde{\theta}^{(t)}$---inherited from the previous optimum $\hat{\theta}^{(t-1)}$---to the optimized state $\hat{\theta}^{(t)}$ by minimizing the recommendation loss $\mathcal{L}_\mathrm{rec}$ (e.g., BPR loss \cite{rendle2009bpr}) on $\mathcal{D}^{(t)}$.

While continual fine-tuning allows adaptation, it suffers from catastrophic forgetting \cite{goodfellow2013empirical, serra2018overcoming}.
To mitigate this, existing works mainly adopt two strategies.
Regularization \cite{wang2023incremental, lee2025leveraging, do2023continual, wang2021graph} imposes constraints to limit deviation from the previous optimum, often by distilling the topological structure of historical interaction graphs \cite{xu2020graphsail, wang2023structure, yoo2025embracing}.
Alternatively, experience replay \cite{lee2024continual, mi2020ader, zhu2023reloop2} maintains a reservoir of representative historical samples for joint training with the current data block, thereby reinforcing past preferences.
However, relying solely on ID-based interaction data limits the efficacy of these strategies in sparse streaming environments, necessitating the integration of LLMs for semantic enhancement.

\subsection{LLM Enhancer for Recommendation} \label{sec/llm_enhancer_for_recommendation}

\begin{figure}[t]
    \centering
    \includegraphics[width=\linewidth]{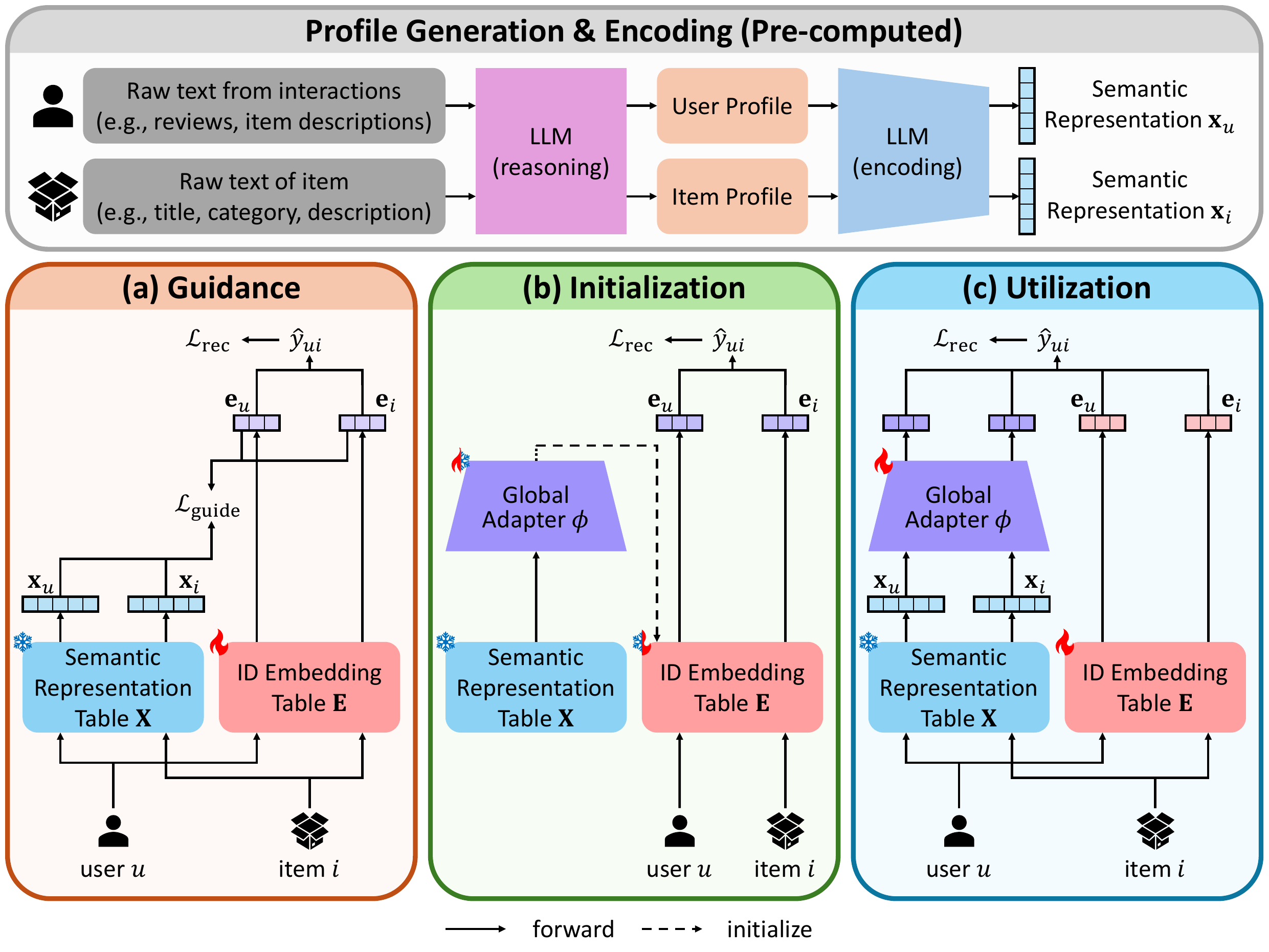}
    \caption{Illustration of three LLM enhancer types.}
    \label{fig/llm_enhancer}
    \Description{}
\end{figure}

\autoref{fig/llm_enhancer} illustrates the LLM enhancer paradigm, augmenting ID-based recommenders with pre-encoded semantic representations to avoid costly direct LLM inference during recommender training and serving.
LLM enhancers often leverage world knowledge and reasoning to transform raw textual metadata into semantically coherent profiles \cite{ren2024representation, xi2024towards}, which are then encoded into $d_\mathrm{LLM}$-dimensional representations and stacked for all users and items to form $\mathbf{X}$, the semantic representation table.\footnote{We focus on the systematic integration of LLM-derived semantic representations into CR. Optimizing the profiling pipeline is beyond the scope of this work.}
In CR, at each stage $t$, new entities ($\mathcal{V}^{(t)}_\mathrm{new}$) are profiled and pre-encoded before model updates.
We classify existing works into three types according to where $\mathbf{X}$ enters the model workflow: \textit{Guidance} in the training objective, \textit{Initialization} at parameter instantiation, and \textit{Utilization} during inference.

\noindent
\textbf{Guidance.}
This strategy \cite{luo2025trawl, liu2024llm, wan2024larr, yang2025darec} aligns evolving ID embeddings with semantic priors via distillation.
In CR, it imposes a semantic constraint to anchor parameters against drift:
\begin{equation}
    \mathcal{L}^{(t)} = \mathcal{L}_\mathrm{rec}(\mathcal{D}^{(t)}) + \lambda_\mathrm{guide}  \sum_{v \in \mathcal{V}^{(t)}} \mathcal{L}_\mathrm{guide}(\mathbf{e}^{(t)}_v, \mathbf{x}_v).
\end{equation}
For $\mathcal{L}_\mathrm{guide}$, RLMRec \cite{ren2024representation} uses contrastive alignment. DLLM2Rec \cite{cui2024distillation} mimics ranking lists inferred by LLMs.
Some methods \cite{li2025ctrl, wang2024flip, wang2025pre} adopt a two-phase strategy: pre-training on $\mathcal{L}_\mathrm{guide}$ and then fine-tuning on $\mathcal{L}_\mathrm{rec}$.

\noindent
\textbf{Initialization.}
This strategy \cite{liu2024llm, hu2024enhancing, zhang2025llminit} projects semantic priors to replace random initialization.
In CR, confining this injection to new entities is desirable to avoid erasing learned history while accelerating adaptation:
\begin{equation}
    \tilde{\mathbf{e}}^{(t)}_v \leftarrow \phi^{(t)}(\mathbf{x}_v), \ \forall v \in \mathcal{V}^{(t)}_\mathrm{new}; \quad \tilde{\mathbf{e}}^{(t)}_v \leftarrow \hat{\mathbf{e}}^{(t-1)}_v, \ \forall v \in \mathcal{V}^{(t)}_\mathrm{ext}.
\end{equation}
For $\phi^{(t)}(\cdot)$, LLM2X \cite{harte2023leveraging} and AlphaFuse \cite{hu2025alphafuse} employ PCA and SVD.
AlphaRec \cite{sheng2025language} and LLMEmb \cite{liu2025llmemb} use MLP adapters, deriving ID embeddings from the adapters rather than maintaining them as independent parameters.

\noindent
\textbf{Utilization.}
This strategy \cite{sun2024large, yang2024sequential, lin2025large, wang2024llm4msr, wan2024larr, luo2025qarm} incorporates semantic representations directly into inference:
\begin{equation}
    \hat{y}_{ui} = f \left( \mathbf{e}^{(t)}_u, \mathbf{e}^{(t)}_i, \phi^{(t)}(\mathbf{x}_u), \phi^{(t)}(\mathbf{x}_i) \right).
\end{equation}
Designs for $\phi^{(t)}(\cdot)$ and $f(\cdot)$ vary significantly.
KAR \cite{xi2024towards} employs MoE for $\phi^{(t)}(\cdot)$ and concatenates $\mathbf{e}^{(t)}$ and $\phi^{(t)}(\mathbf{x})$ within $f(\cdot)$ to form the final representation.
DynLLM \cite{zhao2024dynllm} uses MLP for $\phi^{(t)}(\cdot)$ and cross-attention within $f(\cdot)$ to capture modality interplay.

These strategies are not mutually exclusive and can be combined.
For instance, LLM-ESR \cite{liu2024llm} employs entity-wise $L_2$ regularization for guidance, instantiates $\mathbf{E}$ via PCA projection of $\mathbf{X}$ for initialization, and concatenates $\mathbf{e}$ and $\phi(\mathbf{x})$ to form the final entity representation for utilization.
While these enhancer strategies have proven effective in static settings, extending them to dynamic streaming environments introduces new challenges, particularly in maintaining the delicate balance between stability and plasticity.

\subsection{Problem Definition}

We now formally define our problem.
In LLM-enhanced CR, \textbf{Cognitivity} denotes a model ability to leverage semantic priors grounded in LLM world knowledge and reasoning to infer user-item relevance.
Together with stability for preserving historical preferences and plasticity for adapting to interest shifts, cognitivity forms the following trilemma.

\begin{problem}[Stability-Plasticity-Cognitivity Trilemma]

    \hfill\break\hspace*{1em}\textbf{Given:}
        (1) a model pretrained on a base dataset $\mathcal{D}^{(0)}$;
        (2) a continuous data stream $\{\mathcal{D}^{(t)}\}_{t=1}^{T}$;
        (3) semantic representations $\mathbf{X}$.
    
    \noindent\hspace*{1em}\textbf{Objective:}
        update model parameters $\theta^{(t)}$ at each stage $t$ to balance stability, plasticity, and cognitivity, which are inherently conflicting lemmas in LLM-enhanced CR.

\end{problem}

\section{SPC Trilemma Analysis of LLM Enhancer} \label{sec/spc_trilemma_analysis}

To lay the groundwork for TRACER, we present a comprehensive empirical analysis and interpretation of distinct SPC biases induced by different LLM enhancer types in CR.

\subsection{Lemma Evaluation Metric} \label{sec/lemma_evaluation_metric}

\textbf{Stability.}
We evaluate stability, the capacity to preserve learned history against forgetting, via \textit{Backward Transfer (BWT)} \cite{lopez2017gradient}:
\begin{equation}
    \mathrm{BWT} = \frac{1}{T-1} \sum_{t=1}^{T-1} (R_{T,t} - R_{t,t}),
\end{equation}
where $R_{t, k}$ denotes the model performance (e.g., NDCG) on $\mathcal{D}^{(k)}$ after continuously learning from $\mathcal{D}^{(0)}$ to $\mathcal{D}^{(t)}$.
BWT measures deviation from the performance right after learning $\mathcal{D}^{(t)}$, and a negative value explicitly quantifies forgetting under continual learning.

\noindent
\textbf{Plasticity.}
We assess plasticity, the ability to adapt to emerging distributions, via \textit{Transigence (TRG)}, inspired by \textit{Intransigence} \cite{chaudhry2018riemannian}:
\begin{equation}
    \mathrm{TRG} = \frac{1}{T} \sum_{t=1}^T (R_{t,t} - R_t^*),
\end{equation}
where $R_t^*$ denotes the performance of a model trained from scratch solely on $\mathcal{D}^{(t)}$.
TRG quantifies the gain of continual learning over scratch training, and a negative value implies that past knowledge obstructs current adaptation.

\noindent
\textbf{Cognitivity.}
We evaluate cognitivity, a model ability to leverage semantic priors grounded in LLM world knowledge and reasoning to infer user-item relevance, via a task-based measure.
One could assess this ability by measuring how closely user-item similarity relations in the ID embedding space match those in the semantic representation space.
However, such a representation-level criterion favors semantic mimicry over useful semantic reasoning for recommendation, potentially sacrificing collaborative personalization.
We therefore propose \textit{Semantic Relevance Transfer (SRT)}, which evaluates recommendation performance for new users on tail items under strictly limited adaptation:
\begin{equation}
    \mathrm{SRT} = \frac{1}{T-1} \sum_{t=2}^T R_{t-1, t}^{(k)},
\end{equation}
where $R_{t-1,t}^{(k)}$ denotes the performance of a model, trained up to $\mathcal{D}^{(t-1)}$, on the unseen $\mathcal{D}^{(t)}$ after minimal adaptation using $k$ interactions as training data and $k$ gradient updates.\footnote{We set $k = 5$.}
This setting isolates semantic reliance by suppressing common shortcuts.
New users reduce reliance on accumulated collaborative histories, tail items reduce popularity-driven prediction, and limited adaptation minimizes fresh collaborative evidence.
To implement this evaluation, the support set $\mathcal{S}^{(t)}$ provides $k$ interactions per new user used for adaptation, and the query set $\mathcal{Q}^{(t)}$ is used for evaluation:
\begin{equation}
    \begin{aligned}
        \mathcal{S}^{(t)} & = \bigcup_{u \in \mathcal{U}^{(t)}_\mathrm{new}} \mathrm{Sample} \left( \left\{ (u, i) \in \mathcal{D}^{(t)} \mid i \in \mathcal{I}^{(t)}_\mathrm{ext} \right\}, k \right), \\
        \mathcal{Q}^{(t)}  & = \left\{ (u, i) \in \mathcal{D}^{(t)} \mid u \in \mathcal{U}^{(t)}_\mathrm{new}, i \in \mathcal{I}^{(t)}_\mathrm{ext} \cap \mathcal{I}^{(t)}_\mathrm{tail} \right\},
    \end{aligned}
\end{equation}
where $\mathcal{I}^{(t)}_\mathrm{tail} \subset \mathcal{I}^{(t)}$ denotes tail items (bottom 80\% frequency).
Under this setting, SRT serves as a targeted proxy for a model ability to leverage semantic priors for user-item relevance estimation.

\begin{figure}[t]
    \centering
    \includegraphics[width=\linewidth]{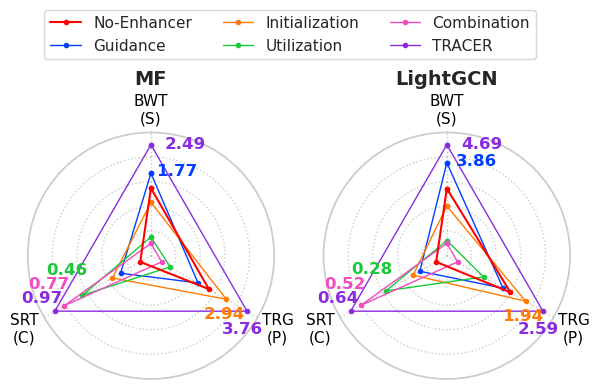}
    \caption{SPC trilemma tendencies of TRACER and LLM enhancers. Annotated numbers (\%) indicate the actual scores of the strongest lemma for each baseline.}
    \label{fig/trilemma_analysis}
    \Description{}
\end{figure}

Using these metrics, we compare \textit{No-Enhancer} (plain ID-based model) with four LLM-enhanced CR baselines, i.e., \textit{Guidance}, \textit{Initialization}, \textit{Utilization}, and their naive \textit{Combination}, alongside our TRACER as a reference, as visualized in \autoref{fig/trilemma_analysis} (see Sec.~\ref{sec/experimental_setup} for baseline details).\footnote{Unless otherwise specified, all reported results in this work are averaged or aggregated across all data blocks and applied continual learning strategies on LightGCN backbone, while consistent trends are observed across all experimental settings.}
The results reveal a structured correspondence: \textit{Guidance}, \textit{Initialization}, and \textit{Utilization} respectively bias LLM-enhanced CR model toward stability, plasticity, and cognitivity, while naive \textit{Combination} fails to harmonize them.
Subsequent sections investigate these SPC biases by formulating and verifying hypotheses to elucidate the underlying phenomena.

\subsection{Guidance: Bias towards Stability}

\textbf{Mechanism.}
Guidance, implemented via RLMRec \cite{ren2024representation}, achieves the highest stability among enhancer baselines but lags behind No-Enhancer in plasticity.
We attribute this to the \textit{semantic anchor} effect, where the guidance loss anchors ID embeddings to semantic representations derived from the invariant world knowledge of LLMs.
This continuously constrains ID embeddings toward fixed semantic positions, acting as a strong regularizer \cite{hinton2015distilling, hsieh2023distilling, xu2025slmrec, wang2024can, kim2026flame, kim2026god}.
While this anchoring effectively filters collaborative noise (e.g., noisy interactions) to preserve core historical preferences, it simultaneously imposes excessive rigidity, obstructing flexible adaptation to evolving distributions \cite{chaudhry2018riemannian}.

\begin{figure}[t]
  \centering
  \begin{subfigure}[t]{0.26\textwidth}
    \centering
    \includegraphics[width=\textwidth]{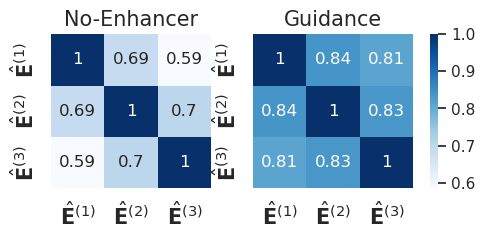}
    \caption{CKA}
  \end{subfigure}
  \begin{subfigure}[t]{0.20\textwidth}
    \centering
    \includegraphics[width=\textwidth]{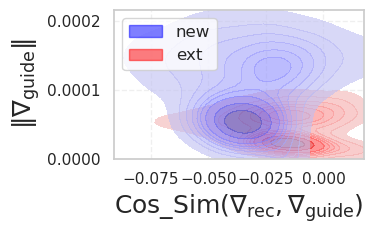}
    \caption{Gradient Distribution}
  \end{subfigure}
  \caption{Guidance analysis. (a) CKA comparing $\hat{\bf E}^{(t)}$ across stage $t$. (b) Gradient distributions of ${\bf E}^{(t)}_\mathrm{new}$ and ${\bf E}^{(t)}_\mathrm{ext}$.}
  \label{fig/guidance_analysis}
  \Description{}
\end{figure}

\noindent
\textbf{Empirical Evidence.}
First, we adopt centered kernel alignment (CKA) \cite{kornblith2019similarity} in \autoref{fig/guidance_analysis}a to measure the structural similarity of optimized ID embeddings across stages.
For each stage pair $(t, k)$, CKA is computed over common entities, $\mathcal{V}^{(t)} \cap \mathcal{V}^{(k)}$.
Guidance maintains higher similarity than No-Enhancer with slower temporal decay, confirming that the semantic anchor suppresses structural shift of ID embeddings (stability $\uparrow$).
Second, gradient dynamics in \autoref{fig/guidance_analysis}b reveal negative cosine similarity between gradients of recommendation ($\nabla_\mathrm{rec}$) and guidance ($\nabla_\mathrm{guide}$) losses, indicating objective conflict.
Crucially, significantly higher $\| \nabla_\mathrm{guide} \|$ on new entities implies that semantic anchoring can overpower early updates when collaborative evidence is scarce, interfering with fresh collaborative signals (plasticity $\downarrow$).

\subsection{Initialization: Bias towards Plasticity}

\textbf{Mechanism.}
Initialization, implemented via LLM2X \cite{harte2023leveraging}, yields the highest plasticity among enhancers yet compromises stability relative to No-Enhancer.
We attribute this to \textit{misaligned semantic topology}.
ID embeddings of new entities inherit a coherent semantic blueprint, which accelerates adaptation by providing informative initial positions \cite{erhan2010does, hou2022towards}.
However, this semantic topology can be misaligned with the collaborative subspace formed by ID embeddings of existing entities, which have been fine-tuned on historical interaction signals \cite{bao2023tallrec, zhu2024collaborative, zhang2025collm}.
Unlike random initialization, which introduces local uncertainty for new entities, semantic initialization injects a structured geometry whose alignment requires global realignment, thereby disrupting existing collaborative structures.

\begin{figure}[t]
  \centering
  \begin{subfigure}[t]{0.23\textwidth}
    \centering
    \includegraphics[width=\textwidth]{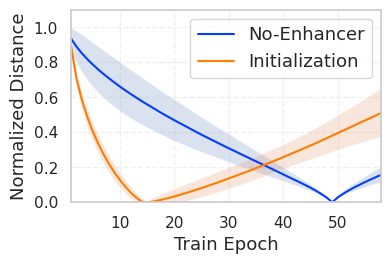}
    \caption{Convergence Trajectory}
  \end{subfigure}
  \begin{subfigure}[t]{0.23\textwidth}
    \centering
    \includegraphics[width=\textwidth]{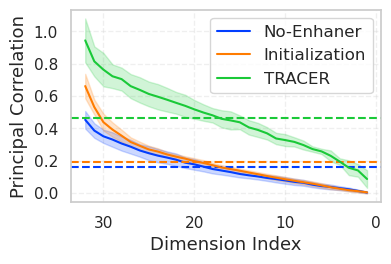}
    \caption{Principal Correlation}
  \end{subfigure}
  \caption{Initialization analysis. (a) Convergence trajectory of ${\bf E}^{(t)}_\mathrm{new}$. (b) Principal correlation between $\tilde{{\bf E}}^{(t)}_\mathrm{new}$ and $\tilde{{\bf E}}^{(t)}_\mathrm{ext}$. Dashed lines show the averages.}
  \label{fig/initialization_analysis}
  \Description{}
\end{figure}

\noindent
\textbf{Empirical Evidence.}
First, \autoref{fig/initialization_analysis}a tracks the distance between new entity ID embeddings $\mathbf{E}^{(t)}_\mathrm{new}$ and their convergence point $\hat{\mathbf{E}}^{(t)}_\mathrm{new}$.
Initialization shows a sharp, low-variance descent compared to No-Enhancer, confirming that the semantic blueprint provides informative initial positions for new entities (plasticity $\uparrow$).
Second, \autoref{fig/initialization_analysis}b employs principal correlation \cite{golub2013matrix} to assess geometric alignment between newly initialized ID embeddings $\tilde{\mathbf{E}}^{(t)}_\mathrm{new}$ and previously learned existing entity ID embeddings $\tilde{\mathbf{E}}^{(t)}_\mathrm{ext}$.
Initialization exhibits low alignment comparable to No-Enhancer, suggesting that the semantic topology is nearly as misaligned with the collaborative subspace as random parameters.
Crucially, unlike unstructured random parameters that can be locally assimilated, fitting this coherent yet misaligned semantic topology requires global realignment, thereby disrupting learned collaborative structures (stability $\downarrow$).

\subsection{Utilization: Bias towards Cognitivity}

\textbf{Mechanism.}
Utilization, implemented via KAR \cite{xi2024towards}, improves cognitivity while severely degrading stability and plasticity.
We attribute this to \textit{semantic dominance} induced by shortcut learning \cite{arpit2017closer, geirhos2020shortcut, shah2020pitfalls, pezeshki2021gradient, kim2026overlooked}.
Since semantic representations enter prediction directly and already carry meaningful information, the model gravitates toward them over randomly initialized ID embeddings of new entities.
Crucially, this shortcut is optimized through a shared global adapter, bypassing the entity-wise parameter isolation \cite{rusu2016progressive, yoon2018lifelong} inherent to ID embeddings.
As adapter updates jointly alter semantic projections across entities, current-block optimization can perturb historical representations \cite{rebuffi2017learning, serra2018overcoming}, while the coarse shared mapping limits entity-specific adaptation.

\begin{figure}[t]
  \centering
  \begin{subfigure}[t]{0.23\textwidth}
    \centering
    \includegraphics[width=\textwidth]{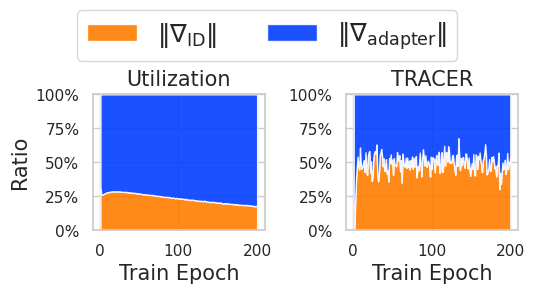}
    \caption{Gradient Magnitude}
  \end{subfigure}
  \begin{subfigure}[t]{0.23\textwidth}
    \centering
    \includegraphics[width=\textwidth]{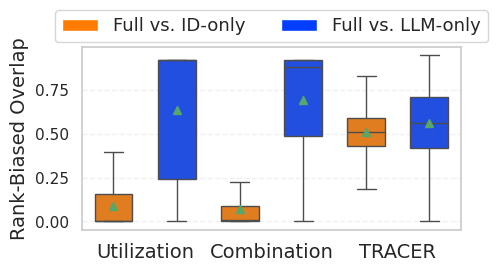}
    \caption{Rank-Biased Overlap}
  \end{subfigure}
  \caption{Utilization analysis. (a) Gradient magnitudes of $\nabla_\mathrm{ID}$ and $\nabla_\mathrm{adapter}$. (b) Rank-biased overlap of full ranking against ID-only and LLM-only rankings.}
  \label{fig/utilization_analysis}
  \Description{}
\end{figure}

\noindent
\textbf{Empirical Evidence.}
First, \autoref{fig/utilization_analysis}a compares gradient magnitudes on ID embeddings ($\nabla_\mathrm{ID}$) and the global adapter ($\nabla_\mathrm{adapter}$).
In Utilization, gradients are concentrated on the adapter, showing that optimization primarily follows the semantic pathway rather than updating ID embeddings.
Second, \autoref{fig/utilization_analysis}b measures rank-biased overlap (RBO) \cite{webber2010similarity} between the final ranking and two single-source rankings, i.e., ID-only rankings from $\mathbf{E}^{(t)}$ and LLM-only rankings from $\phi^{(t)}(\mathbf{X})$.
Utilization closely overlaps with LLM-only rankings but shows negligible overlap with ID-only rankings, indicating over-reliance on semantic knowledge (cognitivity $\uparrow$).
Together, these results confirm shortcut learning in Utilization, where the semantic pathway overshadows collaborative signals and undermines ID-based learning (stability $\downarrow$, plasticity $\downarrow$).

\subsection{Combination: Amplified Cognitivity Bias}
Combination, implemented via LLM-ESR \cite{liu2024llm}, fails to synergistically combine the strengths of Guidance, Initialization, and Utilization.
Despite the highest cognitivity, it exhibits the worst stability and plasticity due to \textit{amplified semantic dominance}.
Unlike standalone Guidance, where fixed semantic representations serve as stable anchors, Combination routes them through a recommendation-trained adapter used for Utilization.
This turns the guidance target into a shifting semantic target, forcing ID embeddings to follow adapter-driven changes and aggravating volatility (stability $\downarrow$).
Meanwhile, continuous guidance locks ID embeddings into this biased semantic topology, negating the adaptability gain of Initialization (plasticity $\downarrow$).
The widened RBO gap in \autoref{fig/utilization_analysis}b relative to Utilization confirms this amplified semantic reliance, indicating that collaborative signals are drowned out (cognitivity $\uparrow$).

\section{Proposed Framework: TRACER}
\begin{figure}[t]
    \centering
    \includegraphics[width=\linewidth]{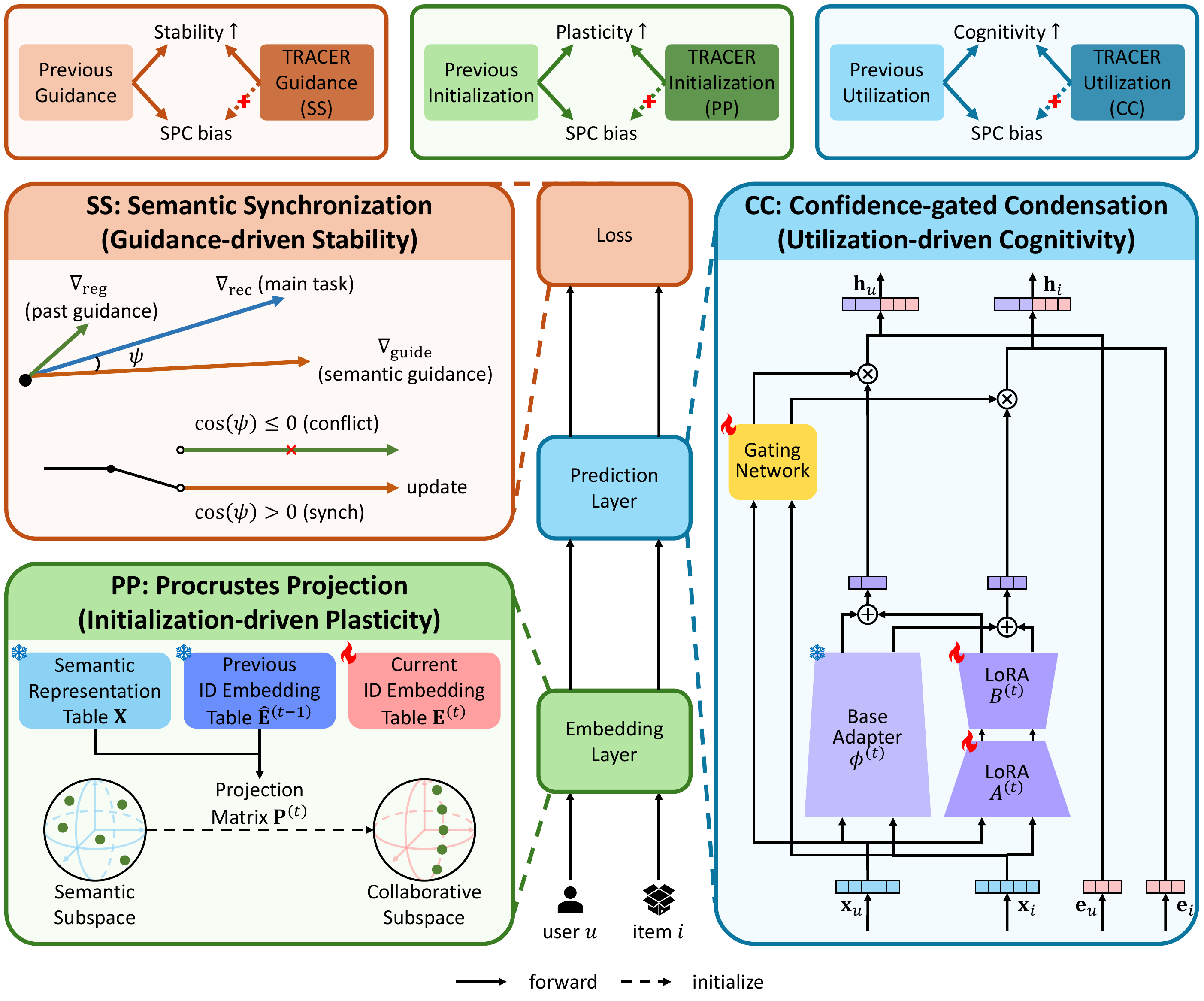}
    \caption{Illustration of TRACER, which redesigns the three LLM enhancer types to preserve their target-lemma strengths while mitigating their inherent SPC biases: \textit{Semantic Synchronization (SS)} for guidance-driven stability, \textit{Procrustes Projection (PP)} for initialization-driven plasticity, and \textit{Confidence-gated Condensation (CC)} for utilization-driven cognitivity.}
    \label{fig/tracer}
    \Description{}
\end{figure}

Building on the analysis in Sec.~\ref{sec/spc_trilemma_analysis}, which revealed that the three enhancer types and their naive combination fail to resolve the SPC trilemma, we propose \textbf{TRACER}, as illustrated in \autoref{fig/tracer}.
TRACER synergistically combines bias-aware redesigns of the three enhancer types, preserving their target-lemma strengths while mitigating inherent SPC biases, thereby achieving superior performance across all lemmas (\autoref{fig/trilemma_analysis}).
Specifically, TRACER aligns the semantic topology with the collaborative subspace for initialization (\autoref{fig/initialization_analysis}b), arbitrates the utilization of dual knowledge sources (\autoref{fig/utilization_analysis}b), and executes guidance that structurally avoids gradient conflicts.
We present the methodology along the learning workflow: (1) instantiating parameters via \textit{initialization}, (2) inferring predictions via \textit{utilization}, and (3) optimizing the objective via \textit{guidance}.

\subsection{Initialization: Procrustes Projection}

Addressing \textit{misaligned semantic topology}, \textbf{Procrustes Projection (PP)} improves plasticity while reducing the stability disruption.
PP performs \textit{topology alignment} by geometrically rotating the semantic subspace toward the collaborative subspace, providing better-aligned positions for new entity ID embeddings.

\paragraph{Global Subspace Alignment}
Unlike prior initialization baselines that ignore the target space, PP computes a projection matrix $\mathbf{P}^{(t)} \in \mathbb{R}^{d_\mathrm{LLM} \times d}$ by aligning semantic representations of previous-stage entities ($\mathcal{V}^{(t-1)}$) with their historical ID embeddings.
We formulate this as the Orthogonal Procrustes problem \cite{schonemann1966generalized}:
\begin{equation}
    \min_\mathbf{P} \| \mathbf{X}_{\mathcal{V}^{(t-1)}} \mathbf{P} - \hat{\mathbf{E}}^{(t-1)} \|_F \quad \text{s.t.} \quad \mathbf{P}^\top \mathbf{P} = \mathbf{I},
\end{equation}
where $\mathbf{X}_{\mathcal{V}^{(t-1)}}$ denotes the rows of $\mathbf{X}$ corresponding to entities in $\mathcal{V}^{(t-1)}$.
The closed-form solution is $\mathbf{P}^{(t)} = \mathbf{V}\mathbf{U}^\top$, via the singular value decomposition of $(\hat{\mathbf{E}}^{(t-1)})^\top \mathbf{X}_{\mathcal{V}^{(t-1)}} = \mathbf{U}\mathbf{\Sigma}\mathbf{V}^\top$.
Crucially, this alignment grounds semantic initialization in the historical coordinate system (stability $\uparrow$).
Simultaneously, discarding the scaling matrix $\mathbf{\Sigma}$ avoids directly importing collaborative scaling patterns (e.g., popularity bias), preserving the distinct semantic geometry.
This orthogonal rotation enables the model to focus on adaptation without correcting subspace mismatch (plasticity $\uparrow$).

\paragraph{Local Calibration}
To resolve local inconsistencies persisting after global alignment, we refine new-entity initialization via entity-wise calibration.
For a new user $u$, we synthesize a collaborative reference $\mathbf{e}_{\mathrm{co}, u}^{(t)}$ by aggregating the semantic projection $\mathbf{e}_{\mathrm{se}, u}^{(t)} = \mathbf{x}_u \mathbf{P}^{(t)}$ with existing-item neighbors $\mathcal{N}(u) = \{ i \in \mathcal{I}^{(t)}_\mathrm{ext} \mid (u, i) \in \mathcal{D}^{(t)} \}$\footnote{For the sake of brevity, we present the formulations for users only, while the operations are applied analogously to items.}:
\begin{equation}
    \begin{aligned}
        \mathbf{e}_{\mathrm{co}, u}^{(t)}   & = \frac{1}{1 + |\mathcal{N}(u)|} \left( \mathbf{e}^{(t)}_{\mathrm{se}, u} + \sum_{i \in \mathcal{N}(u)} \hat{\mathbf{e}}^{(t-1)}_i \right), \\
        \tilde{\mathbf{e}}^{(t)}_u & \leftarrow \beta^{(t)}_u \mathbf{e}_{\mathrm{se}, u}^{(t)} + (1-\beta^{(t)}_u) \mathbf{e}_{\mathrm{co}, u}^{(t)}, & \forall u \in \mathcal{U}^{(t)}_\mathrm{new}
    \end{aligned}
\end{equation}
where $\beta^{(t)}_u = (\mathrm{cos\_sim}(\mathbf{e}^{(t)}_{\mathrm{se}, u}, \mathbf{e}^{(t)}_{\mathrm{co}, u}) + 1) / 2$ down-weights semantic priors that conflict with the collaborative reference (stability $\uparrow$).
This naturally accommodates sparse entities by relying more on the semantic projection $\mathbf{e}_{\mathrm{se}, u}^{(t)}$ when historical support ($|\mathcal{N}(u)|$) is weak (plasticity $\uparrow$).
For pretraining ($t=0$), where historical ID embeddings are unavailable, we use PCA: $\tilde{\mathbf{E}}^{(0)} \leftarrow \mathbf{X} \mathbf{P}_\mathrm{PCA}$ with PCA projection matrix $\mathbf{P}_\mathrm{PCA}$.

\subsection{Utilization: Confidence-gated Condensation} \label{sec/cc}

Mitigating \textit{semantic dominance}, \textbf{Confidence-gated Condensation (CC)} preserves cognitivity while safeguarding stability and plasticity.
CC conducts \textit{semantic modulation}, which controls semantic feature injection via confidence gating and constrains semantic parameter adaptation via low-rank adaptation (LoRA) \cite{hu2022lora}.

\paragraph{Confidence-aware Gating}
We employ a gating network to estimate an injection confidence $c_v^{(t)} \in [0, 1]$ for the adapted semantic representation $\phi^{(t)}(\mathbf{x}_v) \in \mathbb{R}^d$.
We use concatenation to preserve feature independence, so $c_v^{(t)}$ solely controls the semantic injection magnitude.
The final entity representation $\mathbf{h}_v^{(t)}$ is derived as:
\begin{equation}
    c_v^{(t)} = \mathrm{Sigmoid}(W^{(t)}_\mathrm{gate} \mathbf{x}_v + b^{(t)}_\mathrm{gate}), \quad \mathbf{h}_v^{(t)} = [\mathbf{e}_v^{(t)} ; c_v^{(t)} \phi^{(t)}(\mathbf{x}_v)].
\end{equation}
Consequently, the interaction score $\hat{y}_{ui}$ decomposes into ID-based and confidence-weighted semantic terms:
\begin{equation}
    \hat{y}_{ui} = \mathbf{h}^{(t)}_u \cdot \mathbf{h}^{(t)}_i = \mathbf{e}_u^{(t)} \cdot \mathbf{e}_i^{(t)} + c_u^{(t)} c_i^{(t)} \phi^{(t)}(\mathbf{x}_u) \cdot \phi^{(t)}(\mathbf{x}_i),
\end{equation}
where $c_u^{(t)} c_i^{(t)}$ increases semantic intervention only when semantic signals from both user and item are reliable, enabling fine-grained control (plasticity $\uparrow$).
Gating parameters inherit the previous optimum to preserve learned gating criteria across stages ($\tilde{W}^{(t)}_\mathrm{gate} \leftarrow \hat{W}^{(t-1)}_\mathrm{gate}, \tilde{b}^{(t)}_\mathrm{gate} \leftarrow \hat{b}^{(t-1)}_\mathrm{gate}$).

\paragraph{LoRA-based Condensation.}
We apply LoRA to the adapter $\phi^{(t)}(\cdot)$ to restrict the optimization space.
For each layer of $\phi^{(t)}(\cdot)$, we freeze base weights $W^{(t)}_\mathrm{base} \in \mathbb{R}^{d_\mathrm{out} \times d_\mathrm{in}}$ and optimize rank-$r$ matrices $A^{(t)} \in \mathbb{R}^{r \times d_\mathrm{in}}, B^{(t)} \in \mathbb{R}^{d_\mathrm{out} \times r}$ scaled by $\alpha$:
\begin{equation}
    W^{(t)} = W^{(t)}_\mathrm{base} + \frac{\alpha}{r}B^{(t)}A^{(t)}.
\end{equation}
Across stages, we merge the previous residual into the next base weights ($\tilde{W}_\mathrm{base}^{(t)} \leftarrow \hat{W}_\mathrm{base}^{(t-1)} + \frac{\alpha}{r} \hat{B}^{(t-1)} \hat{A}^{(t-1)}$) and re-initialize $A^{(t)}$ and $B^{(t)}$.
Crucially, the bias is fixed to the pretraining optimum ($\tilde{b}^{(t)}_\mathrm{base} \leftarrow \hat{b}^{(0)}_\mathrm{base}$) to prevent severe feature drift.
This confines adaptation to essential low-rank residuals, reducing overfitting (stability $\uparrow$), while the frozen base preserves the semantic core (cognitivity $\uparrow$).
During pretraining ($t = 0$), base parameters ($W^{(0)}_\mathrm{base}, b^{(0)}_\mathrm{base}$) are trained directly without LoRA.

\subsection{Guidance: Semantic Synchronization}

Relaxing the \textit{semantic anchor} effect, \textbf{Semantic Synchronization (SS)} preserves stability while reducing plasticity disruption.
SS turns guidance from a rigid anchor into a \textit{semantic accelerator} by applying semantic guidance only when it is synchronized with the recommendation objective.

\paragraph{Gradient-aware Selective Guidance}
We assess the alignment between collaborative and semantic signals by defining entity-level gradients, $\nabla_{\mathrm{rec}, v} = \nabla_{\mathbf{e}_v^{(t)}} \mathcal{L}_\mathrm{rec}$ and $\nabla_{\mathrm{guide}, v} = \nabla_{\mathbf{e}_v^{(t)}} \mathcal{L}_\mathrm{guide}$ for each entity $v \in \mathcal{V}^{(t)}$.
We quantify this agreement via cosine similarity $\cos(\psi_v) = \mathrm{cos\_sim}(\nabla_{\mathrm{rec}, v}, \nabla_{\mathrm{guide}, v})$.
A positive alignment ($\cos(\psi_v) > 0$) indicates that the semantic guidance supports the current collaborative update.
In this regime, we actively enforce $\mathcal{L}_\mathrm{guide}$ to accelerate convergence (plasticity $\uparrow$).
Conversely, a negative alignment ($\cos(\psi_v) \le 0$) indicates conflicting semantic and collaborative signals.
Here, we deactivate $\mathcal{L}_\mathrm{guide}$ to prevent the semantic anchor from suppressing genuine interest shifts (plasticity $\uparrow$).

\paragraph{Retrospective Regularization}
Deactivating conflicted guidance aids adaptation but risks forgetting existing preferences.
We thus introduce retrospective regularization $\mathcal{L}_\mathrm{reg}$ for existing entities, anchoring each $\mathbf{e}_v^{(t)}$ to its previous optimum $\hat{\mathbf{e}}_v^{(t-1)}$ instead of semantic priors (stability $\uparrow$).
This yields the synchronization objective:
\begin{equation}
    \mathcal{L}_\mathrm{sync}(v) = 
        \begin{cases} 
        \mathcal{L}_\mathrm{guide}(v) & \text{if } \cos(\psi_v) > 0, \\
        \mathcal{L}_\mathrm{reg}(v)   & \text{if } \cos(\psi_v) \le 0 \land v \in \mathcal{V}^{(t)}_\mathrm{ext}, \\
        0                             & \text{otherwise},
        \end{cases}
\end{equation}
where both terms use InfoNCE loss \cite{oord2018representation, liu2021contrastive} with in-batch negatives: $\mathcal{L}_\mathrm{nce}(\mathbf{e}_v^{(t)}, \mathbf{r}_v^{(t)}) = -\log \frac{\exp(\mathbf{e}_v^{(t)} \cdot \mathbf{r}_v^{(t)})}{\sum_{v' \in B} \exp(\mathbf{e}_v^{(t)} \cdot \mathbf{r}^{(t)}_{v'})}$ where $B$ denotes the mini-batch.
Here, $\mathbf{r}_v^{(t)}$ is set to $\phi^{(t)}(\mathbf{x}_v)$ for $\mathcal{L}_\mathrm{guide}$ (using adapter $\phi^{(t)}(\cdot)$ in Sec.~\ref{sec/cc}) and to $\hat{\mathbf{e}}_v^{(t-1)}$ for $\mathcal{L}_\mathrm{reg}$.
The ``otherwise'' case leaves conflicted new entities and pretraining ($t = 0$) optimized only by $\mathcal{L}_\mathrm{rec}$, preventing semantic anchoring from hindering adaptation (plasticity $\uparrow$).
With scaling factor $\lambda_\mathrm{sync}$, the final objective for TRACER at stage $t$ is:
\begin{equation}
    \mathcal{L}^{(t)} = \mathcal{L}_\mathrm{rec}(\mathcal{D}^{(t)}) + \lambda_\mathrm{sync} \sum_{v \in \mathcal{V}^{(t)}} \mathcal{L}_\mathrm{sync}(v).
\end{equation}

\subsection{Synergistic Combination}
The proposed modules establish a virtuous cycle along the learning workflow.
First, PP reduces topological mismatch, allowing CC to gate semantic features based on confidence rather than alignment error, while also reducing gradient conflicts in SS.
CC then curbs semantic dominance, allowing SS to reinforce collaborative updates without rigid semantic anchoring.
Together, these interplays harmonize collaborative and semantic spaces, yielding a high-quality historical optimum $\hat{\mathbf{E}}^{(t)}$ that provides a better reference for future projection $\mathbf{P}^{(t+1)}$.

\section{Experiment}

\subsection{Experimental Setup} \label{sec/experimental_setup}

\subsubsection{Datasets}
\begin{table}[t]
    \scriptsize
    \setlength{\tabcolsep}{3pt}
    \caption{Statistics of the datasets for \autoref{tab/performance}.}
    \label{tab/dataset}
    \begin{tabular}{c|l|c|ccc}
        \toprule
        \multicolumn{2}{c|}{\textbf{Data Blocks}} & $\boldsymbol{\mathcal{D}^{(0)}}$ & $\boldsymbol{\mathcal{D}^{(1)}}$ & $\boldsymbol{\mathcal{D}^{(2)}}$ & $\boldsymbol{\mathcal{D}^{(3)}}$ \\
        \midrule
        \multirow{4}{*}{Home}        & \# Interactions      &  171,457 &   38,100 &   38,100 &   38,100 \\
                                     & \# Accumulated Users &   30,355 &   32,801 &   34,447 &   35,513 \\
                                     & \# Accumulated Items &   16,644 &   17,241 &   17,566 &   17,867 \\
                                     & Sparsity             & 99.97 \% & 99.98 \% & 99.98 \% & 99.97 \% \\
        \hline
        \multirow{4}{*}{CDs}         & \# Interactions      &  299,795 &   66,621 &   66,621 &   66,621 \\
                                     & \# Accumulated Users &   24,633 &   27,589 &   30,752 &   34,485 \\
                                     & \# Accumulated Items &   24,859 &   27,275 &   27,656 &   27,681 \\
                                     & Sparsity             & 99.95 \% & 99.97 \% & 99.97 \% & 99.96 \% \\
        \hline
        \multirow{4}{*}{Movies}      & \# Interactions      &  406,479 &   90,328 &   90,328 &   90,328 \\
                                     & \# Accumulated Users &   27,628 &   34,019 &   42,635 &   46,183 \\
                                     & \# Accumulated Items &   21,612 &   21,872 &   21,917 &   21,920 \\
                                     & Sparsity             & 99.93 \% & 99.97 \% & 99.97 \% & 99.97 \% \\
        \hline
        \multirow{4}{*}{Electronics} & \# Interactions      &  426,769 &   94,837 &   94,837 &   94,837 \\
                                     & \# Accumulated Users &   74,474 &   80,565 &   84,507 &   86,815 \\
                                     & \# Accumulated Items &   31,972 &   34,028 &   35,324 &   35,779 \\
                                     & Sparsity             & 99.98 \% & 99.99 \% & 99.99 \% & 99.99 \% \\
        \hline
        \multirow{4}{*}{Yelp}        & \# Interactions      &  335,385 &   74,529 &   74,529 &   74,529 \\
                                     & \# Accumulated Users &   38,723 &   41,643 &   43,942 &   45,304 \\
                                     & \# Accumulated Items &   26,247 &   27,202 &   27,783 &   28,091 \\
                                     & Sparsity             & 99.97 \% & 99.98 \% & 99.98 \% & 99.98 \% \\
        \bottomrule
    \end{tabular}
\end{table}

\begin{table*}[t]
    \scriptsize
    \setlength{\tabcolsep}{2.9pt}
    \caption{Overall performance (\%). Best and second-best are marked in bold and underlined. Asterisk (*) denotes statistical significance at $p < 0.05$ based on a paired t-test over five independent runs against the strongest baselines.}
    \label{tab/performance}
    \begin{tabular}{c|c|c|cccc|cccc|cccc|cccc|cccc||c|c}
        \toprule
        \multirow{2}{*}{\textbf{Backbone}} & \multirow{2}{*}{\textbf{Metric}} & \multirow{2}{*}{\textbf{Dataset}} & \multicolumn{4}{c|}{\textbf{No-Enhancer}} & \multicolumn{4}{c|}{\textbf{RLMRec (guidance)}} & \multicolumn{4}{c|}{\textbf{LLM2X (initialization)}} & \multicolumn{4}{c|}{\textbf{KAR (utilization)}} & \multicolumn{4}{c||}{\textbf{LLM-ESR (combination)}} & \multirow{2}{*}{\textbf{TRACER}} & \multirow{2}{*}{$\boldsymbol{Improv.}$} \\
        \hhline{~|~|~|----|----|----|----|----||~|~}
        & & & \textbf{FB} & \textbf{FT} & \textbf{RL2} & \textbf{PS} & \textbf{FB} & \textbf{FT} & \textbf{RL2} & \textbf{PS} & \textbf{FB} & \textbf{FT} & \textbf{RL2} & \textbf{PS} & \textbf{FB} & \textbf{FT} & \textbf{RL2} & \textbf{PS} & \textbf{FB} & \textbf{FT} & \textbf{RL2} & \textbf{PS} & & \\
        \midrule\midrule
        \multirow{20}{*}{MF}       & \multirow{4}{*}{ACC}           & Home        &  0.90 &  1.81 &  1.77 &  1.74 &  1.26 &  2.00 &  1.98 &  1.95 &  2.67 &  2.70 &  2.64 &  2.67 &  2.33 &  3.76 &  3.62 &  3.61 &  2.66 &  4.15 &  \underline{4.20} &  4.15 &  \textbf{4.49*} &  6.90 \% \\
                                   & \multirow{4}{*}{(Performance)} & CDs         &  3.98 &  3.37 &  3.38 &  3.27 &  4.13 &  3.48 &  3.50 &  3.46 &  4.80 &  3.82 &  3.81 &  3.84 &  4.95 &  4.87 &  4.81 &  4.95 &  5.58 &  5.56 &  5.42 &  \underline{5.66} &  \textbf{6.06*} &  7.07 \% \\
                                   &                                & Movies      &  3.25 &  2.59 &  2.56 &  2.35 &  3.61 &  2.72 &  2.74 &  2.56 &  4.46 &  3.09 &  3.10 &  2.90 &  4.63 &  5.01 &  4.95 &  4.94 &  5.02 &  5.43 &  5.42 &  \underline{5.44} &  \textbf{5.76*} &  5.88 \% \\
                                   &                                & Electronics &  1.06 &  1.49 &  1.45 &  1.43 &  1.94 &  2.74 &  2.81 &  2.57 &  2.99 &  2.68 &  2.74 &  2.56 &  2.89 &  5.38 &  5.37 &  5.43 &  3.32 &  \underline{5.80} &  5.61 &  5.67 &  \textbf{6.01*} &  3.62 \% \\
                                   &                                & Yelp        &  3.62 &  3.26 &  3.23 &  3.32 &  3.90 &  3.05 &  3.02 &  3.31 &  4.40 &  3.28 &  3.30 &  3.35 &  5.63 &  5.50 &  5.65 &  5.78 &  5.36 &  5.54 &  5.54 &  \underline{6.18} &  \textbf{6.54*} &  5.83 \% \\
        \cline{2-25}
                                   & \multirow{4}{*}{BWT}           & Home        &  0.68 &  0.34 &  0.54 &  0.66 &  \underline{0.97} &  0.58 &  0.75 &  0.77 &  0.61 &  0.31 &  0.40 &  0.58 &  0.32 & -0.03 & -0.04 & -0.13 &  0.28 & -0.11 & -0.08 & -0.16 &  \textbf{1.05*} &  8.25 \% \\
                                   & \multirow{4}{*}{(Stability)}   & CDs         &  1.35 &  0.37 &  0.92 &  0.81 &  \underline{1.68} &  0.74 &  1.27 &  0.66 &  1.30 &  0.33 &  0.69 &  0.71 &  0.44 &  0.08 & -0.06 &  0.09 &  0.38 & -0.57 & -0.56 & -0.71 &  \textbf{1.84*} &  9.52 \% \\
                                   &                                & Movies      &  1.40 &  0.59 &  0.57 &  1.17 &  \underline{1.64} &  0.71 &  0.86 &  1.26 &  1.35 &  0.55 &  0.55 &  0.87 &  0.33 &  0.09 &  0.12 &  0.13 &  0.29 & -0.04 &  0.01 &  0.07 &  \textbf{1.68*} &  2.44 \% \\
                                   &                                & Electronics &  1.59 &  1.15 &  1.18 &  1.24 &  \underline{2.16} &  1.29 &  1.32 &  1.42 &  1.48 &  1.02 &  1.10 &  1.10 &  0.51 & -0.01 &  0.04 &  0.10 &  0.42 & -0.10 & -0.24 &  0.09 &  \textbf{2.20*} &  1.85 \% \\
                                   &                                & Yelp        &  0.51 &  0.39 &  0.41 &  0.44 &  \underline{0.69} &  0.49 &  0.54 &  0.60 &  0.46 &  0.32 &  0.31 &  0.35 &  0.17 & -0.04 & -0.06 & -0.03 &  0.13 & -0.11 & -0.06 & -0.15 &  \textbf{0.70*} &  1.45 \% \\
        \cline{2-25}
                                   & \multirow{4}{*}{TRG}           & Home        &  0.57 &  0.69 &  0.58 &  0.63 &  0.49 &  0.59 &  0.50 &  0.52 &  0.85 &  \underline{1.05} &  0.92 &  0.98 &  0.29 &  0.31 &  0.33 &  0.37 &  0.23 &  0.26 &  0.24 &  0.29 &  \textbf{1.31*} & 24.76 \% \\
                                   & \multirow{4}{*}{(Plasticity)}  & CDs         &  1.38 &  1.52 &  1.52 &  1.41 &  0.91 &  1.27 &  1.07 &  1.11 &  1.95 &  2.14 &  2.08 &  \underline{2.20} &  0.33 &  0.46 &  0.34 &  0.55 & -0.10 &  0.06 &  0.02 &  0.36 &  \textbf{2.63*} & 19.55 \% \\
                                   &                                & Movies      &  1.25 &  1.44 &  1.38 &  1.56 &  0.79 &  1.10 &  0.97 &  1.23 &  1.58 &  \underline{2.05} &  1.90 &  1.99 & -0.02 &  0.37 &  0.18 &  0.39 & -0.50 & -0.04 & -0.07 &  0.09 &  \textbf{2.31*} & 12.68 \% \\
                                   &                                & Electronics &  1.12 &  1.38 &  1.26 &  1.36 &  1.14 &  1.39 &  1.18 &  1.36 &  1.44 &  \underline{1.79} &  1.20 &  1.47 & -0.23 &  0.31 &  0.31 &  0.50 & -0.25 &  0.26 & -0.07 &  0.32 &  \textbf{2.16*} & 20.67 \% \\
                                   &                                & Yelp        &  1.92 &  2.14 &  2.15 &  2.16 &  1.48 &  1.85 &  1.82 &  1.86 &  2.58 &  2.88 &  3.13 &  \underline{3.16} &  1.03 &  0.93 &  1.01 &  1.05 &  0.68 &  0.83 &  0.83 &  1.05 &  \textbf{3.58*} & 13.29 \% \\
        \cline{2-25}
                                   & \multirow{4}{*}{SRT}           & Home        &  0.00 &  0.04 &  0.08 &  0.06 &  0.18 &  0.20 &  0.14 &  0.12 &  0.18 &  0.14 &  0.14 &  0.14 &  0.32 &  0.20 &  0.34 &  0.28 &  0.49 &  \underline{0.65} &  0.40 &  0.52 &  \textbf{0.69*} &  6.15 \% \\
                                   & \multirow{4}{*}{(Cognitivity)} & CDs         &  0.02 &  0.00 &  0.02 &  0.02 &  0.03 &  0.05 &  0.04 &  0.03 &  0.06 &  0.06 &  0.04 &  0.05 &  0.40 &  0.32 &  0.45 &  0.40 &  0.52 &  0.46 &  \underline{0.59} &  0.44 &  \textbf{0.60*} &  1.69 \% \\
                                   &                                & Movies      &  0.02 &  0.02 &  0.02 &  0.03 &  0.04 &  0.03 &  0.03 &  0.04 &  0.04 &  0.06 &  0.10 &  0.14 &  0.18 &  0.18 &  0.26 &  0.22 &  0.22 &  0.28 &  0.26 &  \underline{0.37} &  \textbf{0.38*} &  2.70 \% \\
                                   &                                & Electronics &  0.00 &  0.00 &  0.00 &  0.00 &  0.00 &  0.00 &  0.00 &  0.02 &  0.02 &  0.02 &  0.02 &  0.04 &  0.08 &  0.08 &  0.15 &  0.22 &  0.20 &  0.14 &  0.20 &  \underline{0.24} &  \textbf{0.25*} &  4.17 \% \\
                                   &                                & Yelp        &  0.00 &  0.00 &  0.00 &  0.00 &  0.04 &  0.00 &  0.02 &  0.02 &  0.04 &  0.02 &  0.06 &  0.04 &  0.18 &  0.21 &  0.35 &  0.32 &  0.56 &  0.27 &  0.56 &  \underline{0.68} &  \textbf{0.70*} &  2.94 \% \\
        \midrule
        \multirow{20}{*}{LightGCN} & \multirow{4}{*}{ACC}           & Home        &  1.56 &  2.80 &  2.80 &  2.85 &  1.78 &  2.98 &  2.96 &  3.03 &  2.32 &  3.24 &  3.23 &  3.16 &  2.53 &  4.71 &  4.80 &  4.80 &  2.91 &  5.08 &  5.07 &  \underline{5.36} &  \textbf{5.66*} &  5.60 \% \\
                                   & \multirow{4}{*}{(Performance)} & CDs         &  4.07 &  3.99 &  4.00 &  3.82 &  4.11 &  3.89 &  3.91 &  3.75 &  4.34 &  4.17 &  4.15 &  4.04 &  5.34 &  5.37 &  5.39 &  5.57 &  6.04 &  5.90 &  5.89 &  \underline{6.17} &  \textbf{6.41*} &  3.89 \% \\
                                   &                                & Movies      &  4.02 &  3.01 &  3.02 &  2.84 &  3.99 &  2.95 &  2.97 &  2.76 &  4.46 &  3.24 &  3.27 &  3.11 &  4.72 &  5.16 &  5.06 &  5.14 &  5.39 &  5.74 &  5.73 &  \underline{5.81} &  \textbf{6.22*} &  7.06 \% \\
                                   &                                & Electronics &  2.38 &  2.96 &  2.98 &  2.99 &  2.57 &  3.13 &  3.13 &  3.15 &  3.04 &  3.50 &  3.53 &  3.49 &  3.23 &  5.86 &  5.81 &  5.78 &  3.54 &  6.17 &  \underline{6.20} &  6.19 &  \textbf{6.47*} &  4.35 \% \\
                                   &                                & Yelp        &  4.45 &  3.72 &  3.70 &  3.70 &  4.48 &  3.74 &  3.74 &  3.74 &  4.85 &  3.91 &  3.92 &  3.89 &  5.76 &  6.89 &  6.93 &  7.13 &  5.96 &  \underline{7.30} &  7.19 &  7.26 &  \textbf{8.35*} & 14.38 \% \\
        \cline{2-25}
                                   & \multirow{4}{*}{BWT}           & Home        &  1.48 &  0.93 &  1.36 &  0.94 &  \underline{1.96} &  1.00 &  1.54 &  1.85 &  0.95 &  0.35 &  0.77 &  0.81 &  0.35 & -0.11 & -0.09 & -0.11 &  0.25 & -0.15 &  0.15 &  0.15 &  \textbf{2.12*} &  8.16 \% \\
                                   & \multirow{4}{*}{(Stability)}   & CDs         &  1.92 &  1.08 &  3.35 &  2.03 &  \underline{3.61} &  1.93 &  3.32 &  3.60 &  2.07 &  0.32 &  1.86 &  1.92 &  0.52 &  0.15 &  0.11 &  0.21 &  0.35 &  0.06 &  0.06 &  0.09 &  \textbf{3.67*} &  1.66 \% \\
                                   &                                & Movies      &  1.96 &  1.61 &  2.54 &  1.70 &  \underline{2.67} &  1.69 &  2.49 &  2.50 &  1.98 &  0.54 &  1.70 &  1.76 &  0.71 &  0.08 &  0.07 &  0.04 &  0.54 & -0.06 & -0.08 & -0.05 &  \textbf{2.72*} &  1.87 \% \\
                                   &                                & Electronics &  2.05 &  0.92 &  1.97 &  1.87 &  \underline{2.57} &  2.05 &  2.15 &  2.44 &  1.43 &  0.49 &  1.73 &  1.38 &  0.64 &  0.09 &  0.04 &  0.04 &  0.35 &  0.00 &  0.03 & -0.01 &  \textbf{2.58*} &  0.39 \% \\
                                   &                                & Yelp        &  0.48 &  0.32 &  1.92 &  0.45 &  \underline{2.56} &  0.58 &  2.10 &  2.23 &  0.42 &  0.19 &  0.48 &  0.37 &  0.26 & -0.02 & -0.01 &  0.00 &  0.15 &  0.02 & -0.03 & -0.03 &  \textbf{2.65*} &  3.52 \% \\
        \cline{2-25}
                                   & \multirow{4}{*}{TRG}           & Home        &  0.63 &  0.74 &  0.67 &  1.02 &  0.61 &  0.66 &  0.66 &  0.66 &  0.73 &  \underline{1.11} &  0.92 &  1.04 & -0.58 &  0.33 &  0.29 &  0.37 & -1.51 & -0.59 & -1.39 & -1.05 &  \textbf{1.30*} & 17.12 \% \\
                                   & \multirow{4}{*}{(Plasticity)}  & CDs         &  1.01 &  1.40 &  0.92 &  1.10 &  0.86 &  0.89 &  0.99 &  0.79 &  1.30 &  \underline{2.08} &  1.97 &  1.73 &  0.12 &  0.35 &  0.34 &  0.37 & -0.68 &  0.32 &  0.21 &  0.31 &  \textbf{2.24*} &  7.69 \% \\
                                   &                                & Movies      &  0.65 &  0.59 &  0.56 &  0.63 &  0.29 &  0.37 &  0.34 &  0.29 &  0.54 &  \underline{1.04} &  0.87 &  1.02 &  0.16 &  0.18 &  0.07 &  0.14 & -0.09 &  0.10 &  0.08 &  0.09 &  \textbf{1.26*} & 21.15 \% \\
                                   &                                & Electronics &  0.37 &  0.34 &  0.47 &  0.64 &  0.38 &  0.40 &  0.25 &  0.41 &  0.45 &  0.78 &  \underline{0.93} &  0.84 & -0.19 &  0.28 &  0.22 &  0.30 & -1.99 &  0.08 & -1.50 & -1.12 &  \textbf{1.04*} & 11.83 \% \\
                                   &                                & Yelp        &  0.92 &  1.23 &  0.93 &  1.49 &  0.80 &  1.03 &  0.92 &  0.98 &  1.29 &  1.54 &  \underline{1.56} &  1.48 &  0.38 &  0.43 &  0.39 &  0.40 & -0.26 &  0.37 &  0.11 &  0.40 &  \textbf{2.02*} & 29.49 \% \\
        \cline{2-25}
                                   & \multirow{4}{*}{SRT}           & Home        &  0.16 &  0.16 &  0.17 &  0.20 &  0.14 &  0.14 &  0.14 &  0.16 &  0.15 &  0.14 &  0.14 &  0.14 &  0.28 &  0.19 &  0.20 &  0.32 &  0.48 &  0.47 &  0.43 &  \underline{0.59} &  \textbf{0.60*} & 1.69 \% \\
                                   & \multirow{4}{*}{(Cognitivity)} & CDs         &  0.04 &  0.02 &  0.04 &  0.09 &  0.10 &  0.10 &  0.10 &  0.12 &  0.14 &  0.12 &  0.14 &  0.18 &  0.20 &  0.16 &  0.28 &  0.34 &  0.30 &  0.33 &  0.38 &  \underline{0.39} &  \textbf{0.40*} &  2.56 \% \\
                                   &                                & Movies      &  0.03 &  0.02 &  0.12 &  0.09 &  0.13 &  0.10 &  0.13 &  0.12 &  0.10 &  0.12 &  0.06 &  0.15 &  0.11 &  0.12 &  0.10 &  0.15 &  0.20 &  0.18 &  0.18 &  \underline{0.26} &  \textbf{0.27*} &  3.85 \% \\
                                   &                                & Electronics &  0.00 &  0.00 &  0.00 &  0.02 &  0.00 &  0.00 &  0.00 &  0.00 &  0.00 &  0.00 &  0.00 &  0.00 &  0.02 &  0.02 &  0.02 &  0.04 &  0.13 &  0.08 &  0.13 &  \underline{0.14} &  \textbf{0.15*} &  7.14 \% \\
                                   &                                & Yelp        &  0.00 &  0.00 &  0.00 &  0.00 &  0.00 &  0.00 &  0.00 &  0.00 &  0.02 &  0.00 &  0.00 &  0.02 &  0.08 &  0.13 &  0.18 &  0.18 &  0.12 &  0.21 &  0.19 &  \underline{0.22} &  \textbf{0.26*} & 18.18 \% \\
        \bottomrule
    \end{tabular}
\end{table*}

We evaluate TRACER on four Amazon review categories \cite{mcauley2015image} (\textit{Home, CDs, Movies,} and \textit{Electronics}) and Yelp \cite{asghar2016yelp}.
Following \cite{ren2024representation, lee2024continual}, we apply 10-core filtering and exclude interactions rated below 3 to focus on positive feedback.
To simulate streaming, interactions are ordered chronologically.
The first 60\% form the base block $\mathcal{D}^{(0)}$, and the remaining 40\% are evenly divided into $T = 3$ continual blocks $\{ \mathcal{D}^{(t)} \}_{t=1}^T$.
In each block, interactions for each user are randomly split into train/validation/test sets (8:1:1).
We provide detailed dataset statistics in \autoref{tab/dataset}.

\subsubsection{Baselines}
\begin{itemize}[itemsep=1pt, left=0pt]
    \item \textbf{Backbone}:
        We adopt MF \cite{koren2009matrix} and LightGCN \cite{he2020lightgcn}.
    \item \textbf{Continual Learning}:
        Full-Batch (FB) retrains from scratch on cumulative data $\bigcup_{k=0}^{t} \mathcal{D}^{(k)}$ at stage $t$.
        Fine-Tune (FT) updates the previously learned model using only $\mathcal{D}^{(t)}$.
        ReLoop2 (RL2) \cite{zhu2023reloop2} uses reservoir error memory to calibrate interaction scores.
        PISA (PS) \cite{yoo2025embracing} quantifies user preference shifts for regularization to balance stability and plasticity.
    \item \textbf{LLM Enhancer}:
        No-Enhancer uses only ID embeddings.
        RLMRec \cite{ren2024representation} performs contrastive alignment as guidance.
        LLM2X \cite{harte2023leveraging} employs PCA projection for initialization.
        KAR \cite{xi2024towards} adapts semantic representations with MoE and concatenates them with ID embeddings for utilization.
        LLM-ESR \cite{liu2024llm} combines guidance, initialization, and utilization strategies.
\end{itemize}
For each backbone, we instantiate 20 baselines by pairing four continual learning strategies with five LLM enhancers.

\subsubsection{Evaluation}
We perform full ranking without negative sampling \cite{krichene2020sampled}.
We adopt NDCG@20 as the recommendation metric.
To assess CR, we report average accuracy ($\mathrm{ACC} = \frac{1}{T} \sum_{t=1}^T R_{t, t}$) alongside BWT (stability), TRG (plasticity), and SRT (cognitivity), which are defined in Sec.~\ref{sec/lemma_evaluation_metric}.

\subsubsection{Implementation Details}
We run all models on a single RTX 3090 GPU and average five independent runs.
For semantic representations $\mathbf{X}$, we generate profiles with \texttt{gpt-4o-mini} using the prompt from \cite{ren2024representation} and encode them with \texttt{text-embedding-3-large}.
For LightGCN, we use 3 graph convolution layers.
The ID embedding dimension is $d=64$, except for utilization-based models (KAR, LLM-ESR, TRACER) where $d = 32$ so concatenation with semantic representations yields the same final dimension.
The adapter $\phi^{(t)}(\cdot)$ is a LeakyReLU-activated two-layer MLP projecting $d_\mathrm{LLM} \to (d_\mathrm{LLM} + d) / 2 \to d$.
All models use Adam with learning rate $10^{-3}$, weight decay $10^{-5}$, and batch size 4096, and are trained up to 200 epochs with early stopping patience 30 based on validation ACC.
For TRACER, we grid-search over the synchronization weight $\lambda_\mathrm{sync} \in \{0.01, 0.1, 1.0\}$, LoRA rank $r \in \{4, 8, 16\}$, and LoRA scaling factor $\alpha \in \{8, 16, 32\}$.
Baseline hyperparameters follow the search spaces suggested in their respective papers.

\subsection{Performance Comparison}

\autoref{tab/performance} shows that TRACER consistently outperforms all baselines, achieving up to 14.38\% ACC gain.
TRACER also achieves the best BWT, TRG, and SRT in all cases, demonstrating its superior SPC balance.
Beyond these gains, we provide four key insights:
\begin{itemize}[itemsep=1pt, left=0pt]
    \item
        KAR and LLM-ESR achieve high ACC and SRT as utilization-based enhancers, but show negligible or even negative TRG.
        This suggests that their performance stems from semantic over-reliance rather than effective continual learning dynamics, necessitating systematic integration of LLM enhancers into CR.
    \item
        RLMRec secures high BWT via guidance, while TRACER achieves the highest BWT through SS.
        Crucially, its synergy with PP minimizes topological misalignment, allowing SS to resolve gradient conflicts around genuine interest shifts while preserving history.
    \item
        LLM2X exhibits high TRG via initialization, yet TRACER surpasses this via PP.
        Combined with SS, PP amplifies the semantic accelerator effect by reducing gradient conflicts (\autoref{fig/tracer_ss}a), thereby maximizing responsiveness to emerging patterns.
    \item
        LLM-ESR yields the best baseline ACC but suffers the lowest BWT and TRG despite high SRT, showing that naive combination lacks cohesive module interplay.
        Conversely, TRACER synergistically combines three modules to secure strong SPC scores, ultimately achieving superior ACC.
\end{itemize}

\subsection{Ablation Study}

We assess module contributions using TRACER(PP), TRACER(CC), and TRACER(SS), each retaining only the corresponding module.

\paragraph{PP Analysis}
\begin{figure}[t]
  \centering
  \begin{subfigure}[t]{0.23\textwidth}
    \centering
    \includegraphics[width=\textwidth]{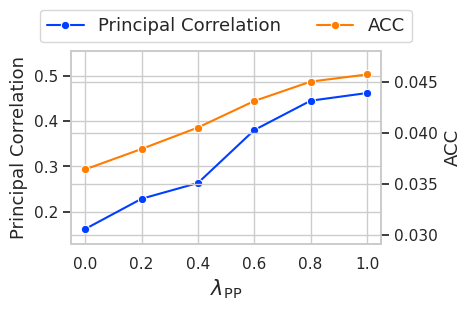}
    \caption{Correlation Sensitivity}
  \end{subfigure}
  \begin{subfigure}[t]{0.23\textwidth}
    \centering
    \includegraphics[width=\textwidth]{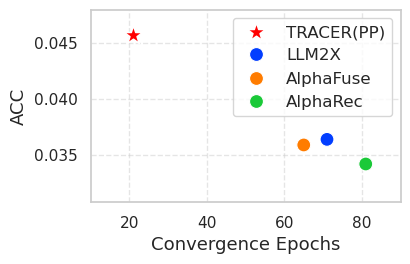}
    \caption{ACC}
  \end{subfigure}
  \caption{TRACER(PP) analysis: (a) Performance and average principal correlation between $\tilde{\mathbf{E}}^{(t)}_\mathrm{new}$ and $\tilde{\mathbf{E}}^{(t)}_\mathrm{ext}$. (b) Performance and convergence epochs of initialization baselines.}
  \label{fig/tracer_pp}
  \Description{}
\end{figure}

To analyze the \textit{topology alignment} effect of PP, we employ geodesic interpolation on the Stiefel manifold \cite{edelman1998geometry}, defining a trajectory $\mathbf{P}(\lambda_\mathrm{PP})$\footnote{$\mathbf{P}(\lambda_\mathrm{PP}) = \mathrm{Exp}^{\mathrm{St}}_{\mathbf{P}_\mathrm{PCA}} \left( \lambda_\mathrm{PP} \mathrm{Log}^{\mathrm{St}}_{\mathbf{P}_\mathrm{PCA}} (\mathbf{P}^{(t)}) \right)$, where $\lambda_\mathrm{PP} \in [0, 1]$. $\mathrm{Exp}^{\mathrm{St}}$ and $\mathrm{Log}^{\mathrm{St}}$ are the Stiefel exponential and logarithm under the canonical metric.} from PCA ($\mathbf{P}(0) = \mathbf{P}_\mathrm{PCA}$) to PP ($\mathbf{P}(1) = \mathbf{P}^{(t)}$).
In \autoref{fig/tracer_pp}a, increasing $\lambda_\mathrm{PP}$ monotonically improves both the principal correlation between $\tilde{\mathbf{E}}^{(t)}_\mathrm{new}$ and $\tilde{\mathbf{E}}^{(t)}_\mathrm{ext}$ and final performance.
This confirms that topology alignment effectively reduces the optimization burden of correcting topology mismatch during training.
We further quantify this benefit in \autoref{fig/tracer_pp}b by comparing TRACER(PP) with initialization baselines (AlphaFuse \cite{hu2025alphafuse}, AlphaRec \cite{sheng2025language}, LLM2X).
TRACER(PP) shows superior convergence and the best performance.
By bypassing the realignment phase, PP allows the model to immediately focus optimization on the core recommendation task.

\paragraph{CC Analysis}
\begin{figure}[t]
  \centering
  \begin{subfigure}[t]{0.27\textwidth}
    \centering
    \includegraphics[width=\textwidth]{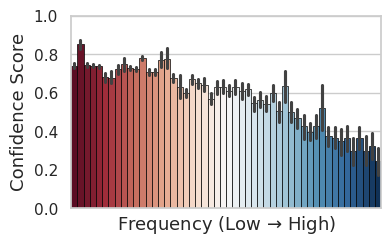}
    \caption{Confidence Score}
  \end{subfigure}
  \begin{subfigure}[t]{0.19\textwidth}
    \centering
    \includegraphics[width=\textwidth]{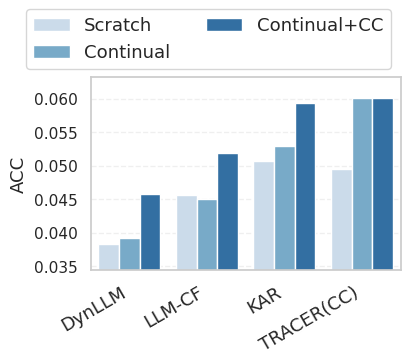}
    \caption{ACC}
  \end{subfigure}
  \caption{TRACER(CC) analysis: (a) Confidence score across entity frequencies. (b) Performance of utilization baselines.}
  \label{fig/tracer_cc}
  \Description{}
\end{figure}

To analyze the \textit{semantic modulation} effect of CC, we examine entity confidence scores $c_v^{(t)}$ across entity frequencies in \autoref{fig/tracer_cc}a.
The inverse correlation between frequency and confidence shows that TRACER(CC) adaptively injects more semantic knowledge into low-frequency entities with scarce collaborative signals, enabling balanced learning.
We further assess CC in \autoref{fig/tracer_cc}b by applying it to utilization baselines (DynLLM \cite{zhao2024dynllm}, LLM-CF \cite{sun2024large}, KAR).
These models suffer from semantic dominance, as continually trained models show negligible gains over scratch-trained counterparts (i.e., low or even negative TRG).
However, CC consistently boosts their performance by adaptively gating semantic injection, alleviating shortcut learning while preserving the stability-plasticity balance of ID embeddings.

\paragraph{SS Analysis}
\begin{figure}[t]
  \centering
  \begin{subfigure}[t]{0.23\textwidth}
    \centering
    \includegraphics[width=\textwidth]{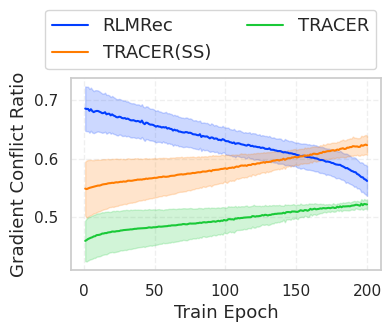}
    \caption{Gradient Conflict}
  \end{subfigure}
  \begin{subfigure}[t]{0.23\textwidth}
    \centering
    \includegraphics[width=\textwidth]{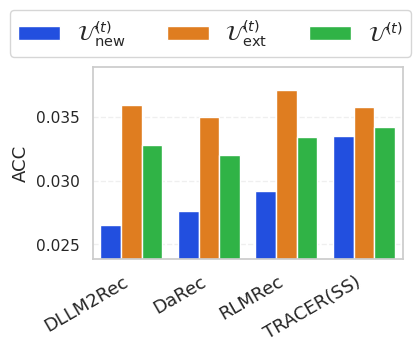}
    \caption{ACC}
  \end{subfigure}
  \caption{TRACER(SS) analysis: (a) Gradient conflict ($\cos(\psi_v) \le 0$) ratio. (b) Performance of guidance baselines.}
  \label{fig/tracer_ss}
  \Description{}
\end{figure}

To verify the \textit{semantic accelerator} effect of SS, we analyze gradient conflicts ($\cos(\psi_v) \le 0$) in \autoref{fig/tracer_ss}a.
We attribute the decreasing ratio of RLMRec to gradual conformity of ID embeddings to a fixed semantic anchor rather than genuine resolution of the objective conflict.
Conversely, the rising ratio under TRACER(SS) may arise as SS blocks conflicting guidance, allowing ID embeddings to accumulate collaborative knowledge and naturally diverge from the semantic anchor.
Full TRACER further lowers conflicts through synergy with PP, which reduces subspace discrepancies.
Compared with guidance baselines (DLLM2Rec \cite{cui2024distillation}, DaRec \cite{yang2025darec}, RLMRec) in \autoref{fig/tracer_ss}b, TRACER(SS) also reduces the performance gap favoring existing users by using semantic guidance to accelerate aligned new-user updates while relying solely on $\mathcal{L}_\mathrm{rec}$ for conflicted updates.

\paragraph{Component Ablation}
\begin{table}[t]
    \scriptsize
    \setlength{\tabcolsep}{3pt}
    \caption{Ablation study with crucial components.}
    \label{tab/ablation_study}
    \begin{tabular}{l|cccc}
        \toprule
        \textbf{Model} & \textbf{ACC} & \textbf{BWT} & \textbf{TRG} & \textbf{SRT} \\
        \midrule
        (1) TRACER                                                        & \textbf{0.0662} &         0.0275  & \textbf{0.0157} & \textbf{0.0034} \\
        \hline
        (2) w/o Retrospective Regularization ($\mathcal{L}_\mathrm{reg}$) &         0.0652  &         0.0148  &         0.0156  &         0.0032  \\
        (3) w/o SS                                                        &         0.0639  &         0.0147  &         0.0144  &         0.0031  \\
        \hline
        (4) w/o Local Calibration ($\mathbf{e}^{(t)}_{\mathrm{co}, v}$)   &         0.0638  &         0.0259  &         0.0143  &         0.0027  \\
        (5) w/o PP                                                        &         0.0608  &         0.0268  &         0.0107  &         0.0029  \\
        \hline
        (6) w/o LoRA-based Condensation ($A^{(t)}, B^{(t)}$)              &         0.0647  &         0.0224  & \textbf{0.0157} &         0.0024  \\
        (7) w/o CC                                                        &         0.0476  & \textbf{0.0276} &         0.0156  &         0.0013  \\
        \bottomrule
    \end{tabular}
\end{table}

To quantitatively validate the effectiveness of components in TRACER, we present the ablation study in \autoref{tab/ablation_study}.
All components improve performance, with three key observations:
\begin{itemize}[itemsep=1pt, left=0pt]
    \item
        The sharp BWT drop in (2) and (3) shows that SS effectively preserves stability.
        The noticeable TRG decline in (3) compared with (2) further suggests that selective guidance successfully mitigates the inherent plasticity drawback of guidance.
    \item
        The low TRG scores of (4) and (5) show that topology alignment in PP is crucial for adaptation.
        The BWT gap between (1) and (4) further indicates that local calibration alleviates the stability deficit of initialization by aggregating historical ID embeddings.
    \item
        The severe SRT drop in (7) confirms that CC is essential for injecting semantic knowledge.
        Meanwhile, the SRT and BWT drops in (6) show that LoRA-based condensation stabilizes semantic adaptation and prevents forgetting.
        Similar BWT and TRG between (1) and (7) indicate that CC improves cognitivity without disrupting the stability-plasticity balance of ID embeddings.
\end{itemize}

\subsection{Practicality Analysis}
\begin{figure}[t]
  \centering
  \begin{subfigure}[t]{0.20\textwidth}
    \centering
    \includegraphics[width=\textwidth]{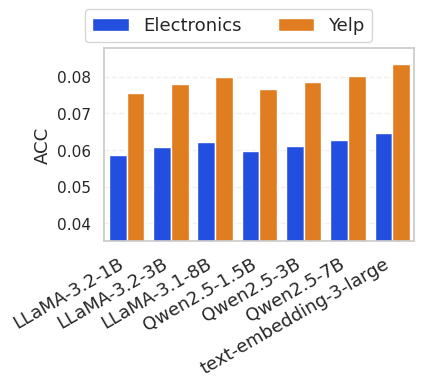}
    \caption{ACC}
  \end{subfigure}
  \begin{subfigure}[t]{0.26\textwidth}
    \centering
    \includegraphics[width=\textwidth]{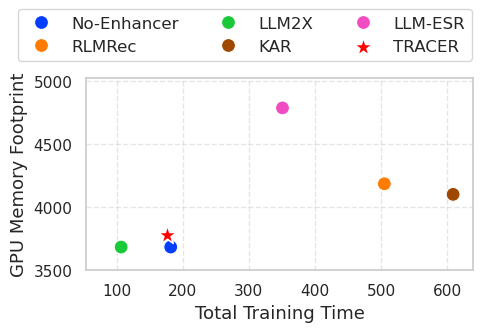}
    \caption{Memory and training time}
  \end{subfigure}
  \caption{(a) Performance of TRACER with different semantic representations encoded via various LLMs. (b) Memory footprint and wall-clock training time of baselines.}
  \label{fig/tracer_practicality_analysis}
  \Description{}
\end{figure}

Beyond accuracy, practical deployment requires robustness to encoder choice, computational efficiency, and low sensitivity to hyperparameters.
We therefore analyze TRACER from these perspectives.

\paragraph{LLM Encoders}
\autoref{fig/tracer_practicality_analysis}a evaluates TRACER across semantic representations from different LLM encoders, using the same profiles generated by \texttt{gpt-4o-mini}.
TRACER performs best with \texttt{text}-\texttt{embedding-3-large} but also remains effective across distinct LLM families (LLaMA \cite{dubey2024llama}, Qwen \cite{qwen2025qwen25technicalreport}).
Notably, performance improves only marginally with larger encoders, suggesting a practical trade-off between LLM encoder size and recommendation accuracy.

\paragraph{Efficiency}
\autoref{fig/tracer_practicality_analysis}b reports GPU memory footprint and wall-clock training time.
TRACER maintains a compact memory footprint close to No-Enhancer, avoiding the high memory overhead of other baselines.
Crucially, TRACER achieves this resource efficiency with rapid convergence, reaching the best performance with minimal computational cost.

\paragraph{Parameter Sensitivity}
\begin{figure}[t]
    \centering
    \includegraphics[width=\linewidth]{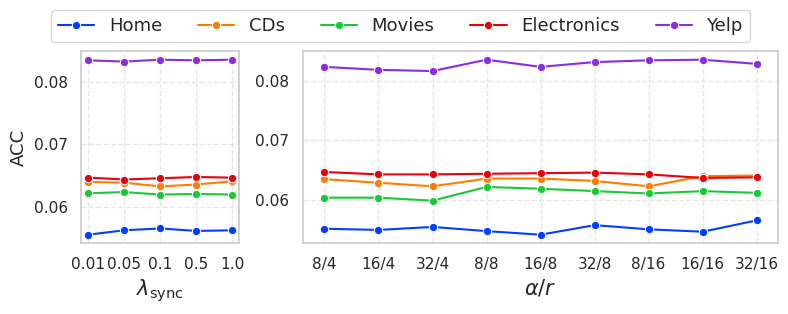}
    \caption{Parameter sensitivity of $\lambda_\mathrm{sync}$, $r$, and $\alpha$.}
    \label{fig/tracer_parameter}
    \Description{}
\end{figure}

\autoref{fig/tracer_parameter} presents the sensitivity analysis for $\lambda_\mathrm{sync}$ in SS and LoRA parameters $r, \alpha$ in CC.
TRACER maintains stable state-of-the-art performance across all configurations, demonstrating its robustness and ease of deployment.

\section{Conclusion}

We identify the SPC trilemma, revealing that naively injecting LLM-derived semantic knowledge into CR disrupts the balance between retaining history and adapting to evolving interests.
To address this, we present TRACER, which synergistically combines three specialized modules to align semantic topology, modulate semantic injection, and resolve gradient conflicts.
TRACER effectively harmonizes stability, plasticity, and cognitivity, achieving state-of-the-art performance and demonstrating that generalized semantic priors can serve as a catalyst for continual adaptation.
Future work includes extending TRACER to multi-modal scenarios by incorporating visual-linguistic models for richer semantic enhancement.

\begin{acks}
    This work was supported by the NRF grant funded by the MSIT (No. RS-2024-00335873), the IITP grant funded by the MSIT (No.RS-2019-II191906, Artificial Intelligence Graduate School Program(POSTECH)).
    This work was also supported by ICT Creative Consilience Program through the IITP grant funded by the MSIT (IITP-2026-RS-2020-II201819) and the IITP grant funded by the MSIT (IITP-2026-RS-2026-25616664, AI Star Fellowship Support Program).
\end{acks}

\appendix
\section{Trilemma Validation Experiment} \label{sec/trilemma_validation_experiment}

\subsection{Datasets}
We utilize four Amazon review categories (\textit{Home, CDs, Movies,} and \textit{Electronics}) and Yelp.
To simulate streaming, interactions are chronologically partitioned into five equal-sized blocks ($\{ \mathcal{D}^{(t)} \}_{t=0}^4$).

\subsection{LLM Enhancer Baselines}

We evaluate canonical baselines for each LLM enhancer type in \autoref{fig/llm_enhancer}, with a scalar $\gamma \in [0, 1]$ to control LLM intervention intensity.

\noindent
\textbf{Guidance}
introduces an auxiliary loss term:
\begin{equation}
    \mathcal{L}_\mathrm{guide}(v) = \| \mathbf{e}^{(t)}_v - \phi^{(t)}(\mathbf{x}_v) \|_2^2,
\end{equation}
where $\mathbf{e}^{(t)}_v$ and $\mathbf{x}_v$ are the ID embedding and semantic representation of entity (i.e., user or item) $v$, and $\phi^{(t)}(\cdot)$ is a two-layer MLP.
With $\mathcal{V}^{(t)}$, the active entities in $\mathcal{D}^{(t)}$, the final objective at stage $t$ is
\begin{equation}
    \mathcal{L}^{(t)} = \mathcal{L}_\mathrm{rec}(\mathcal{D}^{(t)}) + \gamma \sum_{v \in \mathcal{V}^{(t)}} \mathcal{L}_\mathrm{guide}(v).
\end{equation}

\noindent
\textbf{Initialization}
initializes the ID embeddings of $\mathcal{V}^{(t)}_\mathrm{new}$, new entities appearing first in $\mathcal{D}^{(t)}$, using PCA-reduced semantic representations.
Specifically, for a randomly sampled subset $\mathcal{V}^{(t)}_\mathrm{init} \subseteq \mathcal{V}^{(t)}_\mathrm{new}$ of size $|\mathcal{V}^{(t)}_\mathrm{init}| = \lfloor \gamma |\mathcal{V}^{(t)}_\mathrm{new}| \rfloor$, the initialization follows:
\begin{equation}
    \begin{aligned}
            \mathbf{e}^{(t)}_v \leftarrow & \phi_\mathrm{PCA}(\mathbf{x}_v),                & \quad \forall v \in & \mathcal{V}^{(t)}_\mathrm{init}, \\
            \mathbf{e}^{(t)}_v \leftarrow & \mathcal{N}(\mathbf{0}, \sigma^2 \mathbf{I}_d), & \quad \forall v \in & \mathcal{V}^{(t)}_\mathrm{new} \setminus \mathcal{V}^{(t)}_\mathrm{init}.
    \end{aligned}
\end{equation}

\noindent
\textbf{Utilization}
integrates semantic representations during inference.
Specifically, the final entity representation is obtained as follows:
\begin{equation}
    \mathbf{h}^{(t)}_v = [ \mathbf{e}^{(t)}_v; \gamma \phi^{(t)}(\mathbf{x}_v) ].
\end{equation}

\subsection{Evaluation}
Let $R_{t, k}(\gamma)$ denote the NDCG@20 of a model with control scalar $\gamma$ on $\mathcal{D}^{(k)}$ after training from $\mathcal{D}^{(0)}$ to $\mathcal{D}^{(t)}$.
To evaluate stability and plasticity, we measure the performance deviation of the LLM-enhanced CR model relative to the plain ID-based model ($\gamma=0$).

\noindent
\textbf{Stability (x-axis).}
We quantify the change in knowledge retention by averaging the performance differences on all previous blocks:
\begin{equation}
    \frac{1}{t} \sum_{k=0}^{t-1} ( R_{t, k}(\gamma) - R_{t, k}(0) ).
\end{equation}

\noindent
\textbf{Plasticity (y-axis).}
We measure the change in adaptability by calculating the performance difference on the current block:
\begin{equation}
    R_{t, t}(\gamma) - R_{t, t}(0).
\end{equation}

Excluding the first block ($t=0$) where stability cannot be defined, we visualize 20 data points per $\gamma$ value in \autoref{fig/trilemma_validation}a, corresponding to 4 continual steps ($t = 1, 2, 3, 4$) across the five datasets.

\section{GenAI Usage Disclosure}

ChatGPT was used solely for English grammar and language refinement.
No generative AI was involved in the research process, including but not limited to coding, experiments, and data analysis.
All technical contributions are the original work of authors, and AI-assisted edits were manually reviewed to ensure alignment with the original intent.

\clearpage
\bibliographystyle{ACM-Reference-Format}
\balance
\bibliography{references}

\end{document}